\documentclass[11pt]{ip-journal}

\usepackage{macros}
\usepackage[bbgreekl]{mathbbol}
\usepackage{fullpage}

\numberwithin{equation}{section}

\begin{document}

\title{A one-loop exact quantization of Chern-Simons theory}


\date{\today}

\author{Owen Gwilliam}
\address{Department of Mathematics and Statistics \\
Lederle Graduate Research Tower, 1623D \\
University of Massachusetts Amherst \\
710 N. Pleasant Street\\}
\email{gwilliam@math.umass.edu}

\author{Brian Williams}
\address{Department of Mathematics, 
Northeastern University \\ 
567 Lake Hall \\ 
Boston, MA 02115 \\ U.S.A.\\}
\email{br.williams@northeastern.edu}

\maketitle
\thispagestyle{empty}

\begin{abstract}
We examine Chern-Simons theory as a deformation of a 3-dimensional BF theory that is partially holomorphic and partially topological.
In particular, we introduce a novel gauge that leads naturally to a one-loop exact quantization of this BF theory and Chern-Simons theory.
This approach illuminates several important features of Chern-Simons theory,
notably the bulk-boundary correspondence of Chern-Simons theory with chiral WZW theory.
In addition to rigorously constructing the theory,
we also explain how it applies to a large class of closely related 3-dimensional theories 
and some of the consequences for factorization algebras of observables.
\end{abstract}

\tableofcontents

Perturbative Chern-Simons theory admits an exact, one-loop quantization,
and hence the perturbation theory is dramatically simpler than previously realized.
We emphasize that our gauge-fixing respects the three-dimensionality, and it is in the spirit of Axelrod-Singer and Kontsevich, although it lies just outside their framework. 
(Our gauge-fixing is quite different than the axial/lightcone/temporal gauge,
which forces the gauge potential to vanish in one direction, cf.~\cite{FrohKing, MoroSmir}.)
This key result of our paper is a special case of a construction that applies to a large class of theories (and not just in three dimensions),
namely those that are a mixture of topological and holomorphic field theories.
We treat the general situation in a companion paper \cite{GWmixed} but here we will explore the important special case of Chern-Simons theory (and closely related three-dimensional theories).

The idea, which we expand upon below, is that Chern-Simons theory arises as a deformation of the theory that lives on product manifolds $\Sigma \times M$ where $\Sigma$ is a Riemann surface (and hence has a complex structure) and $M$ is a smooth 1-manifold.
Its equations of motion pick out connections that are flat along $M$ and holomorphic along $\Sigma$ (hence {\em mixed} topological-holomorphic).
This theory admits a twist, in the sense of adding a term to the BRST differential, that is usual Chern-Simons theory.
This description also suggests a novel approach to quantizing perturbatively,
by constructing the natural propagator for the mixed theory and then extending to the deformation.

Although this approach may sound obtuse, as it detours through another 3-dimensional theory,
we feel it clarifies some features of Chern-Simons theory.
For instance, this construction relates it to theories that manifestly depend on the complex structure of a Riemann surface.
This approach is perhaps most illuminating when considering Chern-Simons theory on a $3$-manifold with boundary.
There is a natural boundary condition that depends sensitively on the complex structure of the boundary; 
from this perspective it is natural to consider our choice of a propagator, which is defined using the same data.
In this paper we examine in detail the case~$\CC \times \RR_{\geq 0}$. 

The main result of this paper is a construction of a one-loop exact, and finite, quantization of the mixed topological-holomorphic theory on product $3$-manifolds of the form $\Sigma \times M$, which culminates in Theorem~\ref{thm: main}. 
Here, finite refers to the absence of divergences in the Feynman graph expansion.
As a corollary, we recover a one-loop exact quantization of Chern-Simons theory. 
More generally, we characterize a large class of deformations  of the mixed theory that also have such one-loop exact quantizations;
we dub these ``chiral deformations.''
Our result also applies to Chern-Simons-type theories where $\g$ is an $L_\infty$ algebra with nondegenerate, invariant pairing;
this extension encompasses various topological twists of the 3d $\sN = 4$ supersymmetric gauge theories.
With minor modifications to deal with $L_\infty$ spaces, our results should apply to Rozansky-Witten theory and its cousins (cf.~\cite{ChanLeungLi}).

A remarkable feature of this mixed theory is that its quantization is, in fact, a copy of Chern-Simons theory:
the one-loop quantum correction to the action is precisely the deformation of the theory to Chern-Simons theory.
We see this property as giving insight into the special quantization of Chern-Simons theory at the {\em critical level}~\cite{WittenCS, WittenAxelrodPietra}.

In Section~\ref{sec: boundary} we consider the mixed theory on $3$-manifolds with boundary of the form $\Sigma \times \RR_{\geq 0}$,
with a natural ``chiral'' boundary condition.
We show that these mixed theories admit a one-loop exact quantization under this boundary condition.
A striking feature is {\em level-shifting}: the quantization with the boundary condition automatically shifts the level of the classical theory.
Moreover, we explain why our computations should be the key analytical step in phrasing the WZW/CS correspondence using factorization algebras in the style of~\cite{CG1,CG2}.
The level-shifting phenomenon explains how the level used for the chiral WZW theory is related to the level of the bulk Chern-Simons theory.

Our approach also clarifies aspects of the algebras of operators for Chern-Simons theory.
As might be expected due to the manifest appearance of a complex structure,
one obtains a nice approach to canonical quantization of Chern-Simons theory on a Riemann surface,
using the formalism of factorization algebras \cite{CG1,CG2}.
We plan to treat these results in future work.

\subsection{Comparison with other work}

There is an enormous literature on Chern-Simons theory and we make no attempt to survey it here.
Our work falls into the perturbative tradition of Axelrod-Singer \cite{AxeSinI, AxeSinII} and Kontsevich \cite{KonECM},
and it exploits Costello's renormalization machinery \cite{CosBook}, 
which generalized those methods to apply to a broad class of fields theories that arise by adding interaction terms to a free BV theory associated to an elliptic complex.
The idea of understanding gauge theories as deformations of BF theories is well-established \cite{YMasBF},
and one can see its application to Chern-Simons theory in recent work \cite{CMR2,CMW} that is manifestly closely related to the results of this paper.
From the perspective of that recent work, our key innovation is the holomorphic gauge-fixing,
which substantially simplifies the combinatorial complexity of the Feynman diagrams and makes manifest the connection with chiral WZW theories.
It would be interesting to apply our results from the perspective of \cite{MSW},
who offer a different approach to the CS/WZW correspondence.

\subsection{Acknowledgements}

Kevin Costello off-handedly suggested the holomorphic gauge-fixing as a useful approach in the context of Chern--Simons theory,
and we are quite grateful for that comment.
Our Feynman diagrammatic analysis is largely inspired by Si Li's work on the renormalization for chiral conformal field theories \cite{LiFeynman, LiVertex}.
We thank Eugene Rabinovich for extensive discussions on dealing with boundary conditions and for collaborating on verifying Claim~\ref{claim on fact alg} in the abelian case \cite{CSWZW}.
Conversations with Pavel Safronov --- on the CS/WZW correspondence and on shifted Poisson geometry and its role in field theory --- have shaped and aided our thinking;
we thank him and hope his many suggestions will eventually emerge in future work.
The National Science Foundation supported O.G. through DMS Grant No. 1812049.
BW thanks Bruce and Susan Gwilliam for their hospitality during the preparation of this work. 

\section{The mixed classical theory and Chern-Simons theory}

Let us give a gloss of the context of mixed theories.
A classical field theory whose equations of motion have solutions that are locally constant---so that the solutions on a small ball agree with those on a larger ball containing the small ball---are essentially topological in nature.
Classical Chern-Simons theory is of this type, since solutions are flat connections.
It is topological in the sense that the underlying topology of the 3-manifold is all that matters.
A holomorphic classical field theory is one whose equations of motion pick out solutions that only depend on the complex structure of the underlying manifold. 
Classical holomorphic Chern-Simons theory is of this type, since solutions are holomorphic connections on the underlying complex 3-fold.

When the manifold is a product $\Sigma \times M$ of a complex manifold $\Sigma$ and a smooth real manifold $M$, 
it is possible to write down classical theories that are holomorphic along $\Sigma$ and topological along $M$.
(One can sometimes work with manifolds equipped foliations that are transversally holomorphic~\cite{CosVafa}.)
We term such theories {\em mixed},
and a special class of such theories---where $M$ is a 1-manifold---places a crucial role for Chern-Simons theory.

There is an unusual way in which one might have invented Chern-Simons theory as a deformation of such a mixed theory,
and we now describe that artificial history.

\subsection{A false genealogy for Chern-Simons theory}
\label{genealogy for CS}

Consider $\RR^3$ viewed as the product $\CC \times \RR$ with coordinates $(z=x+iy, t)$.
De Rham forms on $\RR^3$ obtain a natural frame in terms of $\dz$, $\dzbar$, and $\dt$,
and the overall de Rham differential decomposes as
\begin{align*}
\d 
&= \del + \delbar + \d_t\\
&= \partial_z \dz + \partial_\zbar + \partial_t \dt\,.
\end{align*}
In brief, there is an isomorphism
\[
\Omega^*(\RR^3) \cong \Omega^{*,*}(\CC) \widehat{\otimes}_\pi \Omega^*(\RR)
\]
of cochain complexes of nuclear spaces, where $\widehat{\otimes}_\pi$ denotes the completed projective tensor product.
What we wish to emphasize here is that forms on $\CC$ admit a double grading as Dolbeault forms.

If we pick a Lie algebra $\g$ with a nondegenerate, invariant bilinear form $\langle-,-\rangle$ (e.g., a semisimple Lie algebra equipped with its Killing form),
then there is a natural mixed gauge theory that we now describe in the BV formalism.

\begin{dfn}
The {\em classical mixed BF theory on $\CC \times \RR$ with values in $\g$} is the following classical BV theory in the sense of~\cite{CosBook}.
The fields consist of the underlying graded vector space of de Rham forms $\Omega^*(\RR^3)[1] \otimes \g$, 
shifted down in cohomological degree.
Let $\alpha$ denote an element of 
\[
\cA = \Omega^{0,*}(\CC) \widehat{\otimes}_\pi \Omega^*(\RR)[1] \otimes \g,
\]
and let $\beta$ denote an element of 
\[
\cB = \Omega^{1,*}(\CC) \widehat{\otimes}_\pi \Omega^*(\RR) \otimes \g.
\]
The action functional is
\[
S_{mix}(\alpha,\beta) = \int \langle \beta, (\delbar + \d_t) \alpha \rangle + \frac{1}{2} \int \langle \beta, [\alpha, \alpha] \rangle,
\]
where we have extended $\langle-,-\rangle$ to a pairing on $\g$-valued de Rham forms with values in de Rham forms.
\end{dfn}

\begin{rmk}
One can also work with a dg Lie or $L_\infty$ algebra $\g$, 
which expands the class of examples considerably.
We do not need $\g$ to be an ordinary Lie algebra above;
an $L_\infty$ algebra with a nondegenerate invariant symmetric pairing of cohomological degree~0 can be used.
From hereon we will work with $\g$ at this high level of generality,
remarking on special features of the case where $\g$ is an ordinary Lie algebra,
particularly when $\g$ is semisimple.
\end{rmk}

The equations of motion are
\[
\delbar \alpha + \d_t \alpha + \frac{1}{2}[\alpha,\alpha] = 0
\]
and
\[
\delbar \beta + \d_t \beta + [\alpha,\beta] = 0.
\]
As usual, let us focus on the degree zero fields and unpack what these equations mean. 
In this case, $\alpha$ is a 1-form, so it has a component of degree $(0,1)$ along $\CC$ and $0$ along $\RR$ and also a component of degree $(0,0)$ and $1$.
Meanwhile, $\beta$ has degree $(1,0)$ along $\CC$ and $0$ along $\RR$.
In other words, we are asking for $\alpha$ such that $\delbar + \d_t + \alpha$ is a connection that is holomorphic along $\CC$ and flat along $\RR$,
and for $\beta$ that is horizontal with respect to this connection and hence a holomorphic 1-form along $\CC$ and constant along $\RR$.
In this sense the classical theory is mixed. 
Some readers will recognize this theory as a BF theory that is holomorphic along $\CC$ and topological along~$\RR$.

\begin{rmk}
Unlike a topological theory, 
which is invariant under all diffeomorphisms of spacetime, 
the action functional of mixed BF theory is naturally invariant under diffeomorphisms 
that are a product of a diffeomorphism in the $\RR$-direction and a biholomorphism in the $\CC$-direction. 
\end{rmk}

It is now straightforward to see how Chern-Simons theory appears as a deformation of this theory. 
Suppose $\g$ is equipped with an invariant, nondegenerate inner product,
and thus fix an isomorphism $\g \cong \g^\vee$.
We now view $B$ as taking values in $\g$.
Now add the interaction term
\[
I_{CS}(\alpha) = \frac{1}{2}\int \langle \alpha, \del \alpha\rangle,
\]
so that $S_{CS} = S_{mix} + I_{CS}$.
The equation of motion for $\alpha$ becomes
\[
\delbar \alpha + \d_t \alpha + \frac{1}{2}[\alpha,\alpha] + \del \alpha = 0,
\]
which is simply the zero curvature equation.

It should be clear that the mixed theory makes sense on any product $\Sigma \times L$ where $\Sigma$ is a Riemann surface and $L$ is a real one-manifold.
The Chern-Simons deformation manifestly extends.

The key observation is this paper can now be stated: 
the mixed theory admits an exact one-loop quantization using the natural gauge-fixing 
(which depends on the complex structure on $\Sigma$),
and this construction extends naturally to the Chern-Simons deformation.
It yields an exact one-loop quantization of Chern-Simons theory.
This feature is well-known for BF-type theories, and we learned it in various works of Cattaneo and Mnev.

\subsection{Some structural features}
\label{sec: structural}

Before we delve into homological algebra in the next section, 
we make some elementary observations about these theories.
With those in sight, the reader may prefer turn to reading about the construction of the quantization,
as the technical homological algebra is essentially independent of the analysis.
Much of the next section is devoted to providing careful formulations of the observations made here.

Consider the following sequence of features:
\begin{itemize}
\item A local functional $I(\alpha,\beta)$ that is purely polynomial (i.e., is only built out of wedging forms and does not involve differential operators) decomposes into a sum of a term $I_0(\alpha)$ that is independent of $\beta$ and a term $I_1(\alpha,\beta)$ that is purely linear in $\beta$. This claim holds because $\beta$ is of $(1,*)$-type in the complex direction.
\item A local functional that only involves {\em holomorphic} derivatives (e.g., $I(\alpha) = \int \alpha \del \alpha$) in the complex direction likewise decomposes into a sum of a term independent of $\beta$ and a term that is linear in~$\beta$.
\item A local functional that involves an antiholomorphic derivative can involve arbitrarily many powers of~$\beta$.
\end{itemize}
Note that the Chern-Simons deformation is of the second kind---involving only a holomorphic derivative---and for this reasons, we will focus in this paper on such {\em chiral} deformations of the mixed BF theory.

\begin{dfn}
A translation-invariant local functional $I(\alpha, \beta)$ is {\em chiral} if its Lagrangian density involves  constant-coefficient differential operators built from $\partial_z$.
We say it has {\em weight zero} if $I$ is independent of $\beta$, and we say it has {\em weight one} if $I$ is linear in~$\beta$.
\end{dfn}

In general, the {\em weight} of a functional will be the number of $\beta$-inputs. 
Below, in Section \ref{sec: moduli} we characterize (up to homotopy) all chiral functionals in mixed BF theory.

\begin{figure}
\begin{tikzpicture}[line width=.2mm, scale = 1]
		\draw[fermion](-5.7,1)--(-5,0);
		\draw[fermion](-4.3,1)--(-5,0);
		\draw[fermion](-5,0)--(-5,-1);
		\filldraw[color=black]  (-5,0) circle (.05);
		\draw (-5.75,1.2) node {$\alpha$};
		\draw (-4.25,1.2) node {$\alpha$};
		\draw (-5,-1.3) node {$\beta$};	
		
		\draw(-5, -2) node{${\rm (a)}$} ;
		
		\draw[fermion] (-1.5, 1)--(-0.5,0);
		\draw[fermion] (-1.5,-1)--(-0.5,0); 
		\draw[fermion] (-0.5,0)--(0.5,0);
		\draw[fermion] (0.5,0)--(1.5,0);
		\draw(0.5,-0.3) node {\small $I_{CS}$};
		\filldraw[color=black]  (-0.5,0) circle (.05); 
		\filldraw[color=black]  (0.5,0) circle (.05);
		\draw (-1.7, 1.2) node {$\alpha$};
		\draw (-1.7, -1.2) node {$\alpha$};
		\draw (-0.3, 0.25) node {$\beta$};
		\draw (0.3, 0.25) node {$\alpha$};
		\draw (0.7, 0.25) node {$\alpha$};
		\draw (1.7, 0) node {$\alpha$};
		\draw(0, -2) node{${\rm (b)}$} ;

		\draw[fermion] (4.3, 1)--(4.8, 0.4);
		\draw[fermion] (4.3, -1)--(4.8, -0.4);
		\draw[fermion] (6.2,0)--(5.5, 0);
		\draw[fill=black] (5,0) circle (.5cm);
		\draw[fill=white] (5,0) circle (.49cm);
		\filldraw[color=black]  (5.5,0) circle (.05);
		\filldraw[color=black]  (4.76,.43) circle (.05);
		\filldraw[color=black]  (4.76,-.43) circle (.05);
		\draw (4.1, 1.3) node {$\alpha$};
		\draw (4.1, -1.3) node {$\alpha$};
		\draw (6.5, 0) node {$\alpha$};
		\draw(5, -2) node{${\rm (c)}$} ;
	    	\clip (0,0) circle (.3cm);

\end{tikzpicture}
\caption{Typical Feynman diagrams. 
The trivalent vertices are labeled by the cubic interaction $\int \langle \beta, [\alpha, \alpha]\rangle$.
The bivalent vertices are labeled by $I_{CS} (\alpha) = \int \langle \alpha, \partial \alpha\rangle$. }
\label{fig: verts}
\end{figure}
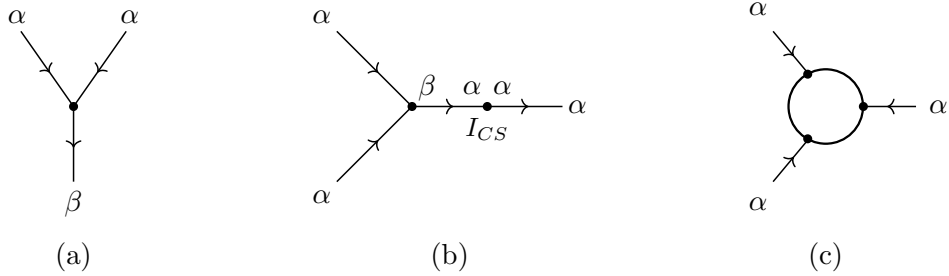

Such chiral functionals behave nicely when constructing Feynman diagrams if we work with a propagator arising from the quadratic functional 
\[
\int \langle \beta, (\delbar + \d_t) \alpha \rangle.
\]
Such a propagator respects the $\alpha$-$\beta$ decomposition of the fields;
graphically, this propagator corresponds to an edge with an $\alpha$-leg and a $\beta$-leg.
Let us view the edge as oriented from the $\alpha$ ``input'' to the $\beta$ ``output.''
A chiral functional has the feature that it has at most one $\beta$-leg,
and so graphically it corresponds to a vertex with at most one outgoing leg.
If one tries to construct graphs using these oriented edges and such vertices, 
one quickly finds that only a small class of connected graphs appear, see Figure \ref{fig: verts}:
\begin{itemize} 
\item[(a)] a tree with a single $\beta$ ``root,''
\item[(b)] a single vertex with no $\beta$-legs, or
\item[(c)] a one-loop graph with no $\beta$-legs (since they get used in making the loop).
\end{itemize}

In other words, we obtain (nonlocal) functionals that are weight zero or weight one.

In anticipation of some results of Section~\ref{sec: hol gauge},
we note that identical observations apply to the anomalies that can appear.
The integral kernel that represents the BV bracket and BV Laplacian respects the $\alpha$-$\beta$ decomposition of the fields, 
just as the propagator does,
so the relevant graphs have the same form (although the analytical content changes). 

We have remarked that the classical action of the mixed BF theory is invariant under the group of diffeomorphisms which are holomorphic in the $\CC$-direction.
In particular, the classical action functional is invariant under the action of $\RR_{>0}$ on $\CC \times \RR$ given by simultaneous dilation of each factor. 
The action $\RR_{>0}$ on fields naturally extends to an action on functionals, in particular on local functionals, by pullback. 

At the level of the quantum theory, the action of $\RR_{>0}$ is known as ``local RG flow.''
In general, a theory that is classically scale-invariant may fail to be scale-invariant at the quantum level, 
which is measured by the $\beta$-function. 
Indeed, the $\beta$-function describes the infinitesimal action by local RG flow in the space of quantum field theories.
For any topological theory, the $\beta$-function is identically zero.
The main analytic result we will prove about mixed BF theory is that no counterterms appear in the Feynman diagram expansion when we employ the holomorphic gauge.
This result means that for mixed BF theory, the $\beta$-function still vanishes. 

The behavior of the local RG flow is related to the BV quantization of the gauge theory. 
Indeed, the vanishing of the $\beta$-function implies that the anomaly to BV quantization is scale-invariant, just as the classical action is. 

\section{Moduli and deformations}
\label{sec: moduli}

We now describe the deformation complex of this theory and identify the Maurer-Cartan element of Chern-Simons theory therein.
This section is not essential for the proof of the one-loop quantization,
and so the reader interested in our key results can skip ahead to the next section.
The results of this section do clarify, however, key structural features of the BV theories in play.

Our primary reference is \cite{CosBook}, notably Chapter 5 on the BV formalism.
The deformation complex is the cochain complex whose underlying graded vector space is the space of local functionals $\oloc$ of the fields. 
The differential is given by the BV bracket against the classical action functional. 

Let $\oloc$ denote the space of local functionals of mixed BF theory on $\CC \times \RR$. 
Then, the deformation complex is given by
\[
\left(\oloc, \{S_{mix},-\}\right)
\]
where $S_{mix}$ is the classical action functional of mixed BF theory. 

First, observe that the differential $\{S_{mix}, -\}$ preserves the weight of a functional,
and hence the deformation complex admits a direct sum decomposition by weight. 

Second, notice that the classical action is translation-invariant. 
We will only consider deformations that are also translation-invariant. 

\begin{dfn}
Let $\Def^{(n)}_{mix}(\fg)$ denote the deformation complex of translation-invariant weight $n$ local functionals for classical mixed BF theory on $\CC \times \RR$ with values in~$\g$.
The full translation-invariant deformation complex (including all weights simultaneously) is the direct sum
\[
\Def_{mix}(\fg) = \bigoplus_{n \geq 0} \Def^{(n)}_{mix} (\fg) .
\]
\end{dfn}

The weight one complex $\Def^{(1)}_{mix}$ encodes deformations that preserve the linearity in the $\beta$-field, 
and hence the role of $\beta$ as a kind of Lagrange multiplier.
(In Section 4.6 of \cite{CG2} classical theories determined by these deformations are called {\em cotangent} theories.)
The weight zero complex $\Def^{(0)}_{mix}$ encodes deformations that are purely functions of the $\alpha$-field,
notably the Chern-Simons deformation.
These are the only two subcomplexes of interest for us,
as we have shown that weight zero and weight one functionals are the only ones that appear in our constructions (both via Feynman diagrams and via chiral deformations).
Further, we have seen that the anomaly to a one-loop quantization must lie in the cohomology of the weight zero part of the deformation complex.

We are interested in a subcomplex of $\Def_{mix}(\fg)$ consisting of deformations that are invariant with respect to a remaining group action. 
We have already mentioned that the classical theory is invariant under all diffeomorphisms along $\RR$ and all holomorphic diffeomorphisms along $\CC$. 
At the linearized level, this means that the classical theory is invariant under the group
\[
{\rm GL}(1 ; \RR) \times {\rm GL}(1 ; \CC) \cong \RR^\times \times \CC^{\times} .
\]
We will restrict ourselves to deformations that are also invariant with respect to this group. 

Here is the main result in the section, which characterizes $\RR^\times \times \CC^\times$-invariant deformations in $\Def_{mix}(\fg)$ of weights zero and one.

\begin{prp}
\label{prop: def}
Let $\fg$ be an $L_\infty$ algebra with nondegenerate invariant pairing of cohomological degree 0. Then
\begin{itemize}
\item[(0)] The $\RR^\times \times \CC^\times$-invariants of the weight zero deformation complex $\Def^{(0)}_{mix} (\fg)$ is quasi-isomorphic to the $2$-term cochain complex
\[
A \overset{\rm def}{=} \left(\C^*_{red}(\fg)[3] \xto{\d} \C^*(\fg, \fg^\vee)[1] \right)
\]
where $\d(\varphi) = 1 \otimes \varphi$ for $\varphi \in \fg^\vee \subset \cO(\fg[1])$, and extends to the graded algebra $\cO(\fg[1])$ by the rule that it is a derivation. 
Explicitly, the quasi-isomorphism is induced by a map of cochain complexes
\[
j^{(0)}  :  A  \to  \Def^{(0)}_{mix}
\]
where $j^{(0)}$ is zero on the component $\C_{red}^*(\fg)[3]$ and for $\mu \in \C^k(\fg ; \fg^\vee)[1]$,
\[
j^{(0)}(\mu) =  \int \ev_{\fg} (\mu(\underbrace{\alpha, \ldots, \alpha}_{k {\rm \; copies}}) , \partial \alpha) ,
\]
where $\ev_g(-,-)$ is the evaluation pairing between $\fg$ and~$\fg^\vee$. 

\item[(1)] The $\RR^\times \times \CC^{\times}$-invariants of the weight one deformation complex $\Def^{(1)}_{mix} (\fg)$ is quasi-isomorphic to $\C^*(\fg ; \fg)$. 
Explicitly, the cochain map is
\[
\begin{array}{ccccc}
j^{(1)} & : & \C^*(\fg ; \fg) & \to & \Def^{(1)}_{mix} \\
& & \xi & \mapsto & \int \langle \beta, \xi(\alpha)\rangle 
\end{array}
\]
is a quasi-isomorphism.
\end{itemize}
\end{prp} 

In the case that $\fg$ is semi-simple, we have the following immediate corollary. 

\begin{crl}\label{crl: def}
Let $\fg$ be a semisimple Lie algebra.
Then
\begin{itemize}
\item[(0)] The cohomology of the $\RR^\times \times \CC^\times$-invariants of the weight zero deformation complex $\Def_{mix}^{(0)}$ is isomorphic to $H^*_{red}(\fg)[3]$. 
Moreover, the image of a degree zero element $\kappa \in H^3(\fg) = \Sym^2(\fg)^\fg$ under the map $j^{(0)}$ is the Chern-Simons deformation $j^{(0)}(\kappa) = \int \kappa(\alpha \partial \alpha)$.
\item[(1)] All $\RR^\times \times \CC^{\times}$-invariant weight one deformations are cohomologically trivial. 
\end{itemize}
\end{crl}

\begin{rmk}
The entire Lie algebra cohomology $H^*(\fg[[z,\varepsilon]])$ has been computed in \cite{FGT}. 
In principle, this could be used to describe the full translation-invariant deformation complex, but it is not obvious how to translate their description to the language of local functionals. 
The description of the invariant piece of the deformation complex we focus on has the benefit of having an explicit description in terms of local functionals. 
\end{rmk} 

 
\subsection{Translation-invariant local functionals}

We will describe the cohomology $\Def_{mix}(\fg)$, in terms of the Lie algebra $\fg$ and its cohomology. 
Our primary tool is Lemma 6.7.1 of \cite{CosBook} which provides a general characterization of the translation-invariant local functionals in any classical theory.
To formulate it, suppose $E$ is the vector bundle on $\CC \times \RR$ underlying the fields of the classical theory.
Then, we can consider the bundle of $\infty$-jets $J E$ and its fiber over $0 \in \CC \times \RR$, $J_0 E$. 
Also note that there is a natural action by the vector fields $\partial_t, \partial_z, \partial_{\zbar}$ on sections of the jet bundle by Lie derivative. 

The translation-invariant deformation complex $\Def_{mix}(\fg)$ can be identified, up to quasi-isomorphism, with the hypercohomology:
\[
{\rm C}_*(\CC^3_{trans} \; ; \; \d t \,\d z\, \d \zbar\, \cO_{red}(J_0 E)) 
\]
Here, $\CC^3_{\rm trans} = \CC\{\partial_t, \partial_z, \partial_{\zbar}\}$ denotes the abelian Lie algebra spanned by the three constant coefficient vector fields, and $\cO_{red}(J_0 E)$ denotes the reduced (quotient by the constant functionals) continuous functionals on $J_0 E$. 
The differential is given by the sum of the Chevalley-Eilenberg differential for $\CC^3_{trans}$ acting on functions on the jet bundle, and the differential $\{S_{mix},-\}$. 

We can identify $J_0 E$, as a graded vector space, with
\[
(\fg[1] \oplus \d z \cdot \fg) \otimes_\CC \CC[[z,\zbar, t]][\d \zbar, \d t] .
\]
The linear part of $\{S_{mix}, -\}$ induces the differential $\delbar + \d_{t}$ on the above graded vector space.
Thus, by the formal Poincar\'{e} lemma, we have a quasi-isomorphism
\[
\left(\cO(J_0 E) , \{S_{mix},-\}\right) \simeq \left(\cO\left( (\fg[1] \oplus \d z \cdot \fg) \otimes_{\CC} \CC[[z]] \right), \{I_{mix},-\}\right)
\]
where $I_{mix}$ is the interacting part of the classical theory. 
Since $I_{mix}$ just encodes the Lie bracket, one can further identify this cochain complex with the following Lie algebra cohomology:
\[
\left(\cO(J_0 E) , \{S_{mix},-\}\right) \simeq \C^*(\fg[[z,\varepsilon]])
\]
where $\varepsilon$ is a formal parameter of degree $+1$.
Note that we have absorbed the factor of $\d z$ into the definition of $\varepsilon$.

In summary, we see that the translation-invariant deformation complex can be identified up to quasi-isomorphism of the following Lie algebra (hyper) homology/cohomology:
\begin{align} \label{deftrans}
\Def_{mix} (\fg) & \simeq \C_*\left(\CC^3_{trans} \; ; \; \d t \,\d z\, \d \zbar\,\C_{red}^*(\fg[[z,\varepsilon]]) \right)
\end{align}

To further simplify the deformation complex, we note that $\C^*(\fg[[z,\varepsilon]])$ does not depend on $t$ or $\zbar$. 
Thus, we have obtained the following. 

\begin{lmm}\label{lmm: def}
There is a quasi-isomorphism
\[
\Def_{mix}(\fg) \simeq \C_*\left(\CC \, \partial_z ; \d z\, \C_{red}^*(\fg[[z,\varepsilon]])\right) \otimes_\CC \d t\, \d \zbar\, \CC[\partial_\zbar, \partial_t]
\]
\end{lmm}

\subsection{Proof of proposition \ref{prop: def}}

Before computing the cohomology, note that the space of $\RR^\times \times \CC^\times$-invariants of $\d t \d \zbar \CC[\partial_\zbar, \partial_t]$ as appearing in Lemma \ref{lmm: def} is one-dimensional spanned by $(\d t \partial_t)(\d \zbar \partial_{\zbar})$ concentrated in degree $-2$. 
Thus, to compute the cohomology of the $\RR^\times \times \CC^\times$-invariant subcomplex, it suffices to compute the hypercohomology
\begin{equation}\label{def2}
\C_*\left(\CC \, \partial_z ; \d z\, \C_{red}^*(\fg[[z,\varepsilon]])\right) [2] .
\end{equation}

We start with the weight zero deformation complex $\Def^{(0)}_{mix}(\fg)$. 
This means we look at the subcomplex that is independent of $\varepsilon$. 
There are two pieces in the deformation complex (\ref{def2}) that are independent of $\varepsilon$ and are $\RR^\times \times \CC^\times$-invariant:
\[
\partial_z \d z \,\C^*_{red}(\fg) [3] \oplus \d z\, \C^*(\fg ; z^\vee \fg^\vee)[1] .
\]
The differential splits into two terms: (1) the internal Chevalley-Eilenberg differential $\d_{CE}$ on each component, and (2) the operator
\[
\frac{\partial}{\partial (\partial_z)} \d : \partial_z \d z\, \C^*_{red}(\fg) [3] \to \d z \C^*(\fg ; z^\vee \fg^\vee)[1]
\]
which is defined on elements $\partial_z \d z\, \varphi \in \partial_z \d z \C^1_{red}(\fg)$ by
\[
 \frac{\partial}{\partial (\partial_z)} \d \left(\partial_z \d z\, \varphi\right) = \d z (1 \otimes \varphi) \in \d z \C^0(\fg; \fg^\vee). 
\]
It extends to a general element in $\partial_z \d z\, \C^*_{red}(\fg)$ by the rule that it is a derivation in the $\C^*_{red}(\fg)$ component. 
This total complex is the complex $A$ in the statement of the proposition. 
One can check that $j^{(0)}$ implements the quasi-isomorphism from $A$ to the $\RR^\times \times \CC^\times$-invariant weight zero deformation complex. 

For the statement about a semi-simple Lie algebra, we compute the cohomology of $A$ by a spectral sequence. 
The first term in the spectral sequence computes the cohomology with respect to the internal Chevalley-Eilenberg differential.  
The second term uses the differential $\frac{\partial}{\partial (\partial_z)} \d$.


We handle the weight one complex $\Def^{(1)}_{mix}$ in a similar way. 
Note that the weight one complex must be linear in $\varepsilon^\vee$. 
Moreover, since $\varepsilon^\vee$ has $\RR^\times \times \CC^\times$ weight $(0,1)$, we see that the $\RR^\times \times \CC^\times$-invariant weight one complex is
\[
\C_*(\CC \, \partial_z \; ; \; \C^*(\fg ; \varepsilon^\vee \fg^\vee)) . 
\] 
Since $\partial_z$ acts trivially on $\C^*(\fg ; \varepsilon^\vee \fg^\vee)$, the same spectral sequence as in the weight zero complex degenerates at the first page, which is simply $H^*(\fg ; \fg^\vee)$.
We can identify this with $H^*(\fg ; \fg)$ via the pairing $\langle-,-\rangle$. 

\section{The holomorphic gauge and associated analysis}
\label{sec: hol gauge}

In this section we analyze the Feynman diagrams that appear in the mixed BF theory and in the relevant deformations, such as Chern-Simons theory.
We begin by constructing the integral kernels for the heat operator and the (mollified) propagator,
and we point out some important properties they possess.
We then examine one-loop diagrams and verify that no counterterms are needed.

Let $V = \CC \times \RR$ and let $\Vpunc$ denote $V \setminus \{0\}$,
the punctured vector space.
We use $(z,t)$ as the linear coordinates on $V$, where $z \in \CC$ is the usual ``holomorphic'' coordinate and $t \in \RR$ in the usual real coordinate.

Let $\Conf_2(V)$ denote $V \times V \setminus {\rm Diag}$, 
where Diag denotes the diagonal copy $V \hookrightarrow V \times V$, 
as it is the space of ordered configurations of two points in~$V$.
We equip it with coordinates $(z,t; w,s)$ where $(z,t)$ are the familiar coordinates on $V$ and $(w,s)$ denote the analogous coordinates on the second copy of~$V$.

There is a natural map $p:  V\times V \to V$ sending $(z,t; w,s)$ to the difference $(z-w,t-s)$.
We will often be interested in the restriction of $p$ to the configuration space,
namely $p: \Conf_2(V) \to \Vpunc$. 
We will find that pullback of forms along $p$ plays a crucial role in analyzing our integral kernels.

\subsection{The gauge fix and associated formulas}

From hereon
\begin{equation}
\label{def of Q}
Q = (\delbar + \d) \otimes \id_\g,
\end{equation}
the term appearing in the quadratic term of the action functional for the mixed BF theory.

\begin{dfn}
The {\em holomorphic gauge fix} is the operator
\begin{equation}
\label{def of Qgf}
Q^* = \left(2\delbar^* + \d^* \right) \otimes \id_\g
\end{equation}
using the Hodge star for the standard Euclidean metric on $\CC \times \RR$.
\end{dfn}

\begin{rmk}
On flat space the explicit formula for $\delbar^*$ and $\d^*$ is
\[
\delbar^* (f \d \zbar) = 2 \frac{\partial f}{\partial z} \;\; , \;\; \d^* (f \d t) = \frac{\partial f}{\partial t} 
\]
where $f \in C^\infty(\CC \times \RR)$. 
\end{rmk} 

\begin{rmk}
On a more general Riemann surface $\Sigma$, we use the terminology of holomorphic gauge fix to mean this operator after a choice of metric on the surface.
\end{rmk}

Ignoring the identity operator on~$\g$,
we see that the Laplacian arising from this gauge fix
\begin{equation}
\label{def of lap}
\triangle = [Q,Q^*] = 2 [\delbar,\delbar^*] + [\d,\d^*] 
\end{equation}
is the standard Laplacian on flat space.

We wish now to construct and analyze the heat kernels and mollified propagators arising from $Q$ and $Q^*$.
It is convenient to introduce the following notation.
These integral kernels are distributional differential forms ({\em aka} currents) on $V \times V$ that are smooth on $\Conf_2(V) = V \times V \setminus {\rm Diag}$.

The heat kernel for this Laplacian {\em acting on functions} is
\[
k^{an}_T(z-w,t-s) = \frac{1}{(4 \pi T)^{3/2}} e^{-|z-w|^2/4T}e^{-|t-s|^2/4T}
\]
at {\em length scale} $T > 0$.
The heat kernel for this Laplacian {\em on the fields} is
\[
K^{an}_T = k^{an}_T  (\dzbar - \d \overline{w}) \wedge (\dz - \d w)\wedge (\d t - \d s).
\]
Note that we have simply multiplied the function $k^{an}_T$ by a differential form,
which is the pullback of a volume form on the diagonal $V$ along the difference map $V \times V \to V$ along the diagonal.
Hence, let's write 
\begin{align*}
\mu_V = \dzbar  \wedge \dz\wedge \d t,
\end{align*}
to introduce some compact notation.
Then $K^{an}_T = p^*( k^{an}_T  \mu_V)$, which is more concise.

We now obtain the mollified propagator
\begin{equation}
\label{def of P}
P_\epsilon^L (z,t; w, s) = \int_{T=\epsilon}^L \d T \, Q^* K_T(z,t; w, s) .
\end{equation}
It's easier to describe this integral kernel by computing $Q^* ( k^{an}_L  \mu_V)$
and then pulling back along $p$.
Explicitly, we find
\begin{equation}
\label{def of diag P}
E_T(z,t) = Q^* (k^{an}_T\mu_V) = -\frac{e^{-(|z|^2 + |t|^2)/4T}}{(4 \pi)^{3/2}T^{5/2}} \dz \wedge \left( \zbar \,\d t +  t \,\dzbar\right).
\end{equation}
In other words, we remove the line element $\dzbar$ or $\dt$ from $\mu_V$ and take the associated derivative of the heat kernel.

It will be more convenient to work with a simpler Gaussian form
\begin{equation}
\label{def of G}
G_T(z,t) = \frac{e^{-(|z|^2 + |t|^2)/4T}}{(4\pi T)^{3/2}} \dz.
\end{equation}
Note that we obtain $E_T$ from $G_T$ by applying a linear combination of the differential operators $\d t \,\partial/\partial \zbar$ and $\dzbar\, \partial/\partial t$:
\begin{equation}
\label{rel of G and E}
E_T(z,t) = \left( \dzbar \frac{\partial}{\partial t} + 2\d t \frac{\partial}{\partial z} \right)G_T(z,t).
\end{equation}
We denote this constant-coefficient differential operator by~$\lambda$.

\subsection{Some computational results}

Later it will be convenient to know how the vector field $\partial/\partial z$ acts on the propagator and the heat kernel.
Direct computation shows that
\begin{equation}
L_{\partial/\partial z} P_\epsilon^L (z,t; w, s) = -\int_{T=\epsilon}^L \d T \, \frac{\zbar-\overline{w}}{4T} E_T(z,t)
\end{equation}
and hence that every further holomorphic derivative multiplies the integrand by a factor $\zbar/4T$.
We note
\begin{equation}
\label{d_z of P}
\frac{\partial^n}{\partial z^n} P_\epsilon^L (z,t; w, s) = -\int_{T=\epsilon}^L \d T \, \frac{(\zbar-\overline{w})^n}{(4T)^n} E_T(z,t)
\end{equation}
and will simply use derivatives (without the Lie derivative notation) from hereon.

We record another convenient computational result.
(It will be used when we study the graph integral of a wheel $\gamma$ with $n+1$ vertices.)
We use coordinates $(q_0,\ldots,q_{n-1}) \in (\CC \times \RR)^{n}$ with $q_i = (z_i,t_i) \in \CC \times \RR$.
Consider the product 
\begin{equation}
\label{prod of Gs}
G_{\vec{T}}^{(n)}(q_0,\ldots,q_{n-1}) = \prod_{i=0}^{n-1} G_{T_i}(q_i) \cdot G_{T_n}(\sum_{i=0}^{n-1} q_i),
\end{equation}
where $\vec{T} = (T_0,\ldots,T_n) \in [\epsilon,L]^{n+1}$.
Direct computation shows the following.

\begin{lmm}
\label{lmm: zeta}
The differential operator
\[
\zeta = \frac{1}{\sum_{j=0}^n T_j} \sum_{i = 0}^{n-1} T_i \frac{\partial}{\partial z_i}
\]
acts on the Gaussian form $G_{\vec{T}}^{(n)}$ by
\[
\zeta(G^{(n)}_{\vec{T}}) = -\frac{\sum_{i = 0}^{n-1} \zbar_i}{4T_n} G^{(n)}_{\vec{T}}.
\]
\end{lmm}

\begin{proof}
Compute that
\begin{equation}
\label{d/dz_jE}
\frac{\partial}{\partial z_i} G^{(n)}_{\vec{T}} = -\frac{1}{4}\left(\frac{\zbar_j}{T_j} + \frac{\sum_{i = 0}^{n-1} \zbar_i}{T_n}\right) E^{(n)}_{\vec{T}}
\end{equation}
by the product rule applied to~\eqref{prod of Gs}. 
Now plug into the definition of~$\zeta$.
\end{proof}

\begin{crl}
\label{clever diff op}
For $0 \leq j < n$, 
\[
\left( \frac{\partial}{\partial z_i} - \zeta\right) G^{(n)}_{\vec{T}} = -\frac{\zbar_j}{4T_j}G^{(n)}_{\vec{T}}.
\]
\end{crl}

For $\partial/\partial t_i$, we have  analogous results.

\begin{lmm}
The differential operator
\[
\tau = \frac{1}{\sum_{j=0}^n T_j} \sum_{i = 0}^{n-1} T_i \frac{\partial}{\partial t_i}
\]
acts on the Gaussian form $G_{\vec{T}}^{(n)}$ by
\[
\tau(G^{(n)}_{\vec{T}}) = -\frac{\sum_{i = 0}^{n-1} t_i}{2T_n} G^{(n)}_{\vec{T}}.
\]
\end{lmm}

\begin{crl}
For $0 \leq j < n$, 
\[
\left( \frac{\partial}{\partial t_i} - \tau\right) G^{(n)}_{\vec{T}} = -\frac{t_j}{2T_j}G^{(n)}_{\vec{T}}.
\]
\end{crl}

We now wish to relate $G^{(n)}_{\vec{T}}$ to the corresponding product of propagators
\begin{equation}
\label{prod of props}
E_{\vec{T}}^{(n)}(q_0,\ldots,q_{n-1}) = \prod_{i=0}^{n-1} E_{T_i}(q_i) \cdot E_{T_n}(\sum_{i=0}^{n-1} q_i).
\end{equation}
By \eqref{rel of G and E}, we see that
\begin{equation}
\label{rel of big G and E}
E_{\vec{T}}^{(n)}(q_0,\ldots,q_{n-1}) = \prod_{i=0}^{n} \lambda_i G^{(n)}_{\vec{T}},
\end{equation}
where
\[
\lambda_i = c_1 \dzbar_i \frac{\partial}{\partial t_i} + c_2 \d t_i \frac{\partial}{\partial z_i}.
\]

Note something important: the differential operators $\lambda_i$, $\zeta$, and $\partial/\partial z_j - \zeta$ mutually commute, as they are constant-coefficient with respect to the $q$ variables.

\subsection{Recollections about Gaussian integrals}
\label{sec on big gaussian}

Finally, consider the Gaussian measure:
\begin{equation}
\label{big gaussian}
\mu(q_0,\ldots,q_n) =  
\exp\left( - \sum_{i=0}^{n-1} \frac{|q_i|^2}{4T_i} - \frac{1}{4T_n}\left|\sum_{i=0}^{n-1} q_i\right|^2\right)
\prod_{i=0}^{n-1}\d q_i.
\end{equation}
Note that its function term and the function term of \eqref{prod of Gs} agree up to an overall factor of $\prod_i T_i^{-3/2}$.
but \eqref{big gaussian} multiplies this function by the Lebesgue measure rather than a purely holomorphic volume form, as with~\eqref{prod of Gs}.

Integrals against this measure, particularly moments, can be quickly computed in terms of the quadratic form appearing in the exponential.
The matrix associated the quadratic form is
\[
M_{\vec{T}} =
\begin{pmatrix}
a_0+b & b & b & \cdots & b \\
b & a_1+b & b & \cdots & b \\
b & b & a_2+b & \cdots & b \\
\vdots & \vdots & \vdots & & \vdots \\
b & b & b & \cdots & a_{n-1}+b
\end{pmatrix}
\]
where $a_i = 1/T_i$ for $0 \leq i \leq n-1$ and $b = 1/T_n$.
As the matrix is the sum of an invertible diagonal matrix $A = (a_i)$ and a rank one matrix $B$ (the all $1/4T_n$ matrix),
the Sherman-Morrison theorem \cite{SM} gives an explicit formula for the inverse matrix:
\[
M_{\vec{T}}^{-1} = \begin{pmatrix}
c_0 + d_{00} & d_{01} & d_{02} & \cdots & d_{0, n-1} \\
d_{10} & c_1+ d_{11} & d_{12} & \cdots & d_{1, n-1} \\
d_{20} & d_{21} & c_2 + d_{22}& \cdots & d_{2,n-1} \\
\vdots & \vdots & \vdots & & \vdots \\
d_{n-1, 0}& d_{n-1, 1} & d_{n-1, n-2}& \cdots & c_{n-1}+ d_{n-1,n-1}
\end{pmatrix}
\]
where 
\[
c_i = T_i  \;\;\; , \;\;\; d_{jk} = d_{kj} = - \frac{T_{j} T_k}{T_0 + \cdots + T_n}
\]
for $0 \leq i ,j , k\leq n-1$. 

The inverse determinant is
\begin{equation}
\label{matrix det}
\det\left( M_{\vec{T}}^{-1}\right) = \frac{T_0 T_1 \cdots T_n}{T_0 + T_1 + \cdots + T_n}
\end{equation}
by the matrix determinant lemma.
This expression makes computing moments straightforward.

\section{There are no divergences}

Our main technical result is the following.

\begin{thm}
Any chiral deformation of mixed BF theory admits a one-loop exact and finite (pre)quantization. 
\end{thm}

The meaning of this theorem is the following. 
For any chiral deformation of the mixed gauge theory on $\CC \times \RR$ with $L_\infty$ algebra $\g$,
the only Feynman diagrams that appear have at most one loop.
This property is what we mean by one-loop exact.
(We already explained why only graphs with at most one loop appear in Section~\ref{sec: structural}.)
The finiteness part of the assertion is that the associated Feynman integrals have no ultraviolet divergences when one uses the holomorphic gauge fix and associated parametrices.
This part of the theorem involves some explicit calculus, which we perform in this section. 
We use the term {\em pre\/}quantization to emphasize that it remains to check the quantum master equation 
(in physical terms, that this one-loop correction is compatible with the path integral measure);
we verify that in the next section.

This theorem follows from a pair of results about one-loop diagrams.

\begin{lmm}
\label{lmm: less than 3}
If a wheel $\gamma$ has one vertex or two vertices, then the analytic weight $w^{an}_\gamma(P_\epsilon^L,I)$ vanishes for all $0 < \epsilon < L < \infty$.
In particular, the $\epsilon \to 0$ limit is zero.
\end{lmm}

\begin{proof}
When the wheel has a single vertex, 
the propagator $P_\epsilon^L$ is evaluated along the diagonal,
where it vanishes. 
Hence the associated graph integral vanishes (even at finite $\epsilon$ and $L$). 

When the wheel has two vertices, the graph integral vanishes for a slightly more subtle reason. 
Before dealing with a general case of an arbitrary chiral interaction,
let us focus on the analysis involving just the internal loop.

Label the vertices by coordinates $(z_1, t_1)$ and $(z_2,t_2)$. 
The loop consists of two edges connecting the vertices, and each edge corresponds to a propagator whose inputs are the coordinates.
As we have observed, the propagator depends only on the difference $(z,t) = (z_1-z_2, t_1 -t_2)$.
As our two edges have the same coordinates, 
we get the square of a single propagator, 
but the term $\dz \wedge \dz$ kills the square.

This argument does not quite suffice, 
because the vertices correspond to local functionals,
which contribute nontrivial factors to the overall integral.
If, however, the functionals are polynomials in $\alpha$ and $\beta$ (i.e., without any derivatives in the Lagrangian density), 
then this argument shows the graph integral vanishes at any~$\epsilon$.

Hence, now consider the case where at least one vertex is labeled by a local functional that involves differential operators built from the $\partial_{t_i}$ and $\partial_{z_i}$,
using the coordinates labeling the vertices.
These can act on the propagators labeling the internal edges, 
by the Lie derivatives $L_{\partial_{t_i}}$ and $L_{\partial_{z_i}}$. 
By formula \eqref{def of diag P}, we see that these Lie derivatives produce integral kernels whose product is zero.
\end{proof}

\begin{prp}
\label{prp: more than 2}
If a wheel $\gamma$ has more than two vertices, then the analytic weight $w^{an}_\gamma(P_\epsilon^L,I)$ has a well-defined $\epsilon \to 0$ limit as a distribution.
\end{prp}

To prove this proposition, we will show that for any compactly-supported input $\phi$,
the absolute value of the weight $|w^{an}_\gamma(P_\epsilon^L,I)(\phi)|$ is bounded above by an integral whose dependence on $\epsilon$ converges as it goes to zero.
To do find this convenient upper bound, 
we first use integration by parts to move all derivatives of the propagators onto $\phi$.
This new integral is manifestly an integral against a Gaussian measure followed by an integral over the ``length scale'' parameters $T_i$,
and so we can bound the integral over the $T_i$ by applying Wick's lemma to the Gaussian measure.

\begin{proof}
Let the wheel have $n+1$ vertices.
To write the integral kernel for $w[\epsilon]$, we label the vertices in consecutive order from $0$ to $n$ as we go around the wheel.
To the $j$th vertex we assign a coordinates $p_j = (w_j,s_j) \in \CC \times \RR$.
To the edge from the $j$th to $(j+1)$st vertex, we assign the propagator $P_\epsilon^L(p_j, p_{j+1})$.
Each vertex is assigned some chiral interaction term,
but its details are not so important.
By definition, it will take an input function $f(p_j)$ and apply some constant-coefficient differential operator built from $\partial/\partial w_j$ to $f$; 
it will apply other such holomorphic differential operators to the incoming propagators;
and then it will multiply these outputs together.
Without loss of generality, we can view the action on $f$ as absorbed into the choice of an input;
what matters is that the propagators may be acted upon by constant-coefficient holomorphic differential operators.

Every graph integral $w^{an}_\gamma(P_\epsilon^L,I)$ thus determines a distribution taking as input any compactly-supported function on $(\CC \times \RR)^{n+1}$.
Its integral kernel is smooth, as we will see.
We will leave $L>0$ fixed throughout, and we will not adjust the interaction $I$. 
For concision we denote the distribution $w[\epsilon]$, as we are interested in its behavior as $\epsilon$ goes to zero.

In the $p$-coordinates we thus find that our integral kernel has the following form
\[
w[\epsilon](\phi) = \int_{\vec{T} \in [\epsilon,L]^{n+1}} \d^{n+1} T \, \int_{\vec{p} \in (\CC \times \RR)^{n+1}} \d^{n+1} p \, \phi(\vec{p}) \prod_{j=0}^{n} D_j E_{T_j}(p_j, p_{j+1})
\]
where $D_j$ is a constant-coefficient differential operator built from $\partial/\partial w_j$ 
and $E_{T_j}$ is the Gaussian-type differential form from~\eqref{def of diag P}.

Each propagator depends only on the difference $q_j = p_{j+1} - p_j$, where $j$ runs from $0$ to $n-1$,
and so it is convenient to rewrite this integral using these coordinates.
We write $q_j = (z_j, t_j) \in \CC \times \RR$ and note that
\[
p_n - p_0 = q_0 + q_1 + \cdots + q_{n-1},
\]
which is relevant to the propagator between the 0th and $n$th vertices.
We use $\d q_j$ to represent the volume form $\dz_j \, \dzbar_j \, \d t_j$.
Set $q_n = p_n$.
In these $q$-coordinates our integral kernel is
\[
w[\epsilon](\phi) = \int_{\vec{T} \in [\epsilon,L]^{n+1}} \d^{n+1} T \, \int_{\vec{q} \in (\CC \times \RR)^{n+1}} \d^{n+1} q \,  \phi(\vec{q}) \prod_{j=0}^{n-1} D'_j E_{T_j}(q_j) \left( D'_n E_{T_n}(\sum_{j=0}^{n-1} q_j)\right)
\]
where the $D'_j$ are the $D_j$ re-expressed in these new $q$ coordinates.
In light of equation~\eqref{d_z of P}, 
we see that the $D'_j$ are simply constant-coefficient differential operators built from~$\partial/\partial z_j$.

Without loss of generality, we may assume that each $D'_j$ is a monomial differential operator of the form $(\partial/\partial z_j)^{k_j}$,
as an arbitrary $D'_j$ can be written as a sum of such.
By equation~\eqref{d_z of P} and Corollary~\ref{clever diff op}, we know that 
\[
\left(\frac{\partial}{\partial z_j}\right)^{k_j} E_{T_j}(q_j)  = \left( -\frac{\zbar_j}{4T_j}\right)^{k_j} E_{T_j}(q_j) = \left(  \frac{\partial}{\partial z_j} - \zeta\right)^{k_j} E_{T_j}(q_j).
\]
We also know from~\eqref{rel of G and E} that
\[
E_{T_j}(q_j) = \lambda_j G_{T_j}(q_j) = \left( c_1 \dzbar \frac{\partial}{\partial t_j} + c_2 \d t \frac{\partial}{\partial z_j} \right) G_{T_j}(q_j)
\]
with $G_T(q)$ from~\eqref{def of G} and $\lambda_j$ a constant-coefficient differential operator.
Hence
\[
D'_j E_{T_j}(q_j)  = \left(  \frac{\partial}{\partial z_j} - \zeta\right)^{k_j} \lambda_j G_{T_j}(q_j).
\]
Note that $\lambda_j$ and $\partial/\partial z_j - \zeta$ commute.

Putting these observations together and ignoring the input term $\phi$, our integrand can be written as
\[
\left(\prod_{i=0}^{n-1} \left(  \frac{\partial}{\partial z_i} - \zeta\right)^{k_i} \right) \zeta^{k_n}\left( \prod_{j=0}^{n-1} \lambda_j \right) G^{(n)}_{\vec{T}}.
\]
In other words, our integrand arises by applying a product of differential operators to $G^{(n)}_{\vec{T}}$.
We denote the full operator by~$D_{\vec{k}}$.

What makes this description convenient is that it allows us to apply integration by parts:
we can move all the differential operators acting on  $G^{(n)}_{\vec{T}}$ over to the input $\phi$.
We thus find that
\[
w[\epsilon](\phi) = \pm\int_{\vec{T} \in [\epsilon,L]^{n+1}} \d^{n+1} T \, \int_{\vec{q} \in (\CC \times \RR)^{n+1}} \d^{n+1} q \, 
\left(D_{\vec{k}} \phi(\vec{q}) \right)
G^{(n)}_{\vec{T}}.
\]
In short, we have reduced our graph integral to computing the expected value of $D_{\vec{k}} \phi(\vec{q})$ against a Gaussian measure depending on $\vec{T}$,
as mentioned in the overview of the proof.

This expression has the drawback that the differential operator $D_{\vec{k}}$ explicitly depends on the length scale parameters $\vec{T}$, 
but the dependence is easy to bound.
Note that for any smooth function $f(q_0,\ldots,q_0)$, 
\[
|\zeta f| 
\leq \frac{1}{\sum_{j=0}^n T_j} \sum_{i = 0}^{n-1} T_i \left| \frac{\partial f}{\partial z_i}\right| 
\leq  \sum_{i = 0}^{n-1} \left| \frac{\partial f}{\partial z_i} \right|,
\]
as $T_i/(T_0 + \cdots + T_n) < 1$ because the $T_j$ are all positive real numbers.
In consequence, $|D_{\vec{k}} \phi(\vec{q},p_n)|$ is bounded by some compactly-supported, nonnegative function $\psi$ on $(\CC \times \RR)^{n+1}$ that is independent of~$\vec{T}$.

In short, we now know that
\[
|w[\epsilon](\phi)| \leq \int_{\vec{T} \in [\epsilon,L]^{n+1}} \d^{n+1} T \, \int_{\vec{q} \in (\CC \times \RR)^{n+1}} \d^{n+1} q \, 
 \psi(\vec{q}) |G^{(n)}_{\vec{T}}(\vec{q})|
\]
Note that $|G^{(n)}_{\vec{T}}(\vec{q})|$ is almost the Gaussian measure
\[
\frac{1}{\prod_i T_i^{\frac{3}{2}}}\exp\left( - \sum_{i=0}^{n-1} \frac{|q_i|^2}{4T_i} - \frac{1}{4T_n}\left|\sum_{i=0}^{n-1} q_i\right|^2\right)
\prod_{i=0}^{n-1}\d q_i,
\]
except that it is missing the $\d t_i$ and $\dzbar_i$ factors of the Lebesgue measure.
These form factors must be supplied by $\psi$ to obtain something integrable, 
which imposes conditions on which kinds of inputs matter.

If $\psi$ contributes the missing factors,
we can bound $|w[\epsilon](\phi)|$ by an integral over $(q_0,\ldots,q_{n-1})$ against the Gaussian measure~\eqref{big gaussian}.
Thus the results of Section \ref{sec on big gaussian} kick in
by Taylor expanding $\psi$ and using standard formulas for moments of Gaussians.
In this case, the zeroth moment contributes a multiple of
\[
\frac{1}{T_0^{3/2} \cdots T_n^{3/2}} \left(\frac{T_0 T_1 \cdots T_n}{T_0 + T_1 + \cdots + T_n}\right)^{3/2} = \frac{1}{(T_0 + T_1 + \cdots + T_n)^{3/2}}.
\]
The integral over the $T_i$ factors then converges in the $\epsilon \to 0$ limit.
To show this more explicitly, we note that the arithmetic-geometric mean inequality tells us that
\[
\frac{T_0 + T_1 + \cdots + T_n}{n+1} \geq (T_0 T_1 \cdots T_n)^{\frac{1}{n+1}},
\]
so that
\begin{align*}
\int_{\vec{T} \in [\epsilon,L]^{n+1}} \frac{\d^{n+1} T}{(T_0 + T_1 + \cdots + T_n)^{3/2}} 
&\leq \int_{\vec{T} \in [\epsilon,L]^{n+1}} \frac{\d^{n+1} T}{(T_0 T_1 \cdots T_n)^{\frac{3}{2(n+1)}}}\\
&= \left(\left( 1- \frac{3}{2(n+1)} \right) \left( \epsilon^{1-\frac{3}{2(n+1)}} - L^{1- \frac{3}{2(n+1)}} \right) \right)^{n+1}.
\end{align*}
Now, we have $n+1 > 2$ by hypothesis, so $3/2(n+1) < 1$ and the bound on the right has a finite $\epsilon \to 0$ limit.

Higher moments for $\psi$ contribute, by Wick's lemma, more terms of the form $T_j T_k/(T_0 + T_1 + \cdots + T_n)$ 
and hence are even better behaved in the $\epsilon \to 0$ limit.
Indeed, the $k$th term in the Taylor expansion of $\psi$ contributes terms of the form
\begin{align*}
\frac{1}{T_0^{3/2} \cdots T_n^{3/2}} & \left(\frac{T_0 T_1 \cdots T_n}{T_0 + T_1 + \cdots + T_n}\right)^{3/2} \left(\frac{T_{i_1} T_{j_1}}{T_0 + \cdots + T_n}\right) \cdots \left(\frac{T_{i_k} T_{j_k}}{T_0 + \cdots + T_n}\right) \\ & =  \frac{1}{(T_0 + T_1 + \cdots + T_n)^{3/2}}\left(\frac{T_{i_1} T_{j_1}}{T_0 + \cdots + T_n}\right) \cdots \left(\frac{T_{i_k} T_{j_k}}{T_0 + \cdots + T_n}\right) .
\end{align*}
where $(i_1, j_1), \ldots, (i_k, j_k)$ are pairs of integers satisfying $0 \leq i_\ell, j_\ell \leq n$. 
For these higher moments, the integral we must bound is of the form
\[
\int_{\vec{T} \in [\epsilon,L]^{n+1}}\frac{\d^{n+1} T}{(T_0 + T_1 + \cdots + T_n)^{3/2}} \left(\frac{T_{i_1} T_{j_1}}{T_0 + \cdots + T_n}\right) \cdots \left(\frac{T_{i_k} T_{j_k}}{T_0 + \cdots + T_n}\right) .
\]
For any of the pairs $(i_{\ell}, j_{\ell})$, we have the inequality
\[
\frac{T_{i_\ell} T_{j_\ell}}{T_0 + \cdots + T_n} \leq T_{i_{\ell}},
\]
since $T_{j_{\ell}} \leq T_0 + \cdots + T_n$.
It follows that we can bound the integral from above by
\[
\int_{\vec{T} \in [\epsilon,L]^{n+1}}\frac{T_{i_1} \cdots T_{i_k}}{(T_0 + T_1 + \cdots + T_n)^{3/2}}\d^{n+1} T,
\]
which is finite in the $\epsilon \to 0$ limit (when $n + 1 > 2$) by applying the AM/GM inequality to the denominator as in the argument for the zeroth moment.
\end{proof}

\begin{rmk}
A reader familiar with \cite{LiFeynman} might have noticed parallels between the analysis performed above and the analogous one-loop analysis in chiral conformal field theory on Riemann surfaces. 
That work did indeed provide a guide.
Based on various work of Li, we thus expect that for classes of partially holomorphic theories on $\RR \times \CC$ that are {\em not} exact at one-loop (say, theories that are not of cotangent type), one can adjust the arguments in \cite{LiFeynman, LiVertex} to show that such theories are finite to {\em all orders in $\hbar$}.
\end{rmk}

\section{BV quantization}

In this section we show that the renormalization defined by the chiral gauge fixing condition provides a solution to the quantum master equation (QME) for mixed BF theory in the presence of an arbitrary chiral deformation. 
In general, there are anomalies to solving the quantum master equation measuring the extent to which a classical gauge symmetry fails to be a symmetry at the quantum level. 
We will find that for analytic reasons, based on our gauge fixing condition introduced above, that such an anomaly in our setting vanishes identically (see Lemma~\ref{lmm: anomaly}). 

This result, combined with our work in the previous section, yields the main result of this paper. 

\begin{thm} 
\label{thm: main}
Any chiral deformation of mixed BF theory on $\CC \times \RR$ admits a one-loop exact quantization. 
\end{thm}

In light of our results on deformations of the mixed theory, 
summarized in Proposition \ref{prop: def} and Corollary \ref{crl: def}, 
we obtain the following corollary. 

\begin{crl} 
When $\fg$ is semi-simple, the space of all translation-invariant, ${\rm GL}_1(\CC) \times {\rm GL}_1(\RR)$-equivariant cotangent quantizations of mixed BF theory on $\CC \times \RR$ is a torsor for $\Sym^2(\fg)^\fg$. 
\end{crl}
 
This corollary about the space of quantizations of mixed BF theory bears a striking resemblance to the quantizations of ordinary Chern-Simons theory. 
In Theorem 14.2.1 of \S5 in \cite{CosBook}, 
it is shown that on any $3$-manifold $M$, 
the space of BV quantizations for Chern-Simons theory is a torsor for the group $\hbar H^3(\fg) [[\hbar]]$,
which is isomorphic to $\hbar \Sym^2(\fg)^\fg [[\hbar]]$ when $\fg$ is semi-simple. 
Thus, to one-loop on $\CC \times \RR$, 
the spaces of quantizations of mixed BF theory and Chern-Simons theory are isomorphic. 

The relation between mixed BF theory and Chern-Simons can be made even more direct.
By our results, a given $\hbar$-dependent level $\hbar \kappa \in \hbar \Sym^2(\fg)^\fg$ corresponds to a quantization of mixed BF theory.
When $\kappa$ is not trivial, and upon inverting $\hbar$, this quantization is isomorphic to the quantization of Chern-Simons theory at level $\kappa$.

One can ask whether there is a distinguished choice of an $\hbar$-dependent level,
as there are many choices.
The chiral gauge picks out a preferred quantization on $\CC \times \RR$ for which the $\hbar$-dependent level is zero. 
This feature changes when one constructs a quantization of mixed BF theory in the presence of boundary conditions. 
In Section~\ref{sec: boundary} we show that for a natural choice of boundary condition for holomorphic BF theory on the $3$-manifold with boundary $\CC \times \RR_{\geq 0}$, 
the quantization determined by the chiral gauge corresponds to a non-trivial choice in level proportional to the Killing form of~$\fg$. 

\subsection{BV structures}

Our approach to quantization is through the Batalin-Vilkovisky (BV) formalism, which classically uses the data of the of the pairing of degree $(-1)$ on the fields.
For mixed BF theory, this pairing utilizes the pairing on $\fg$ together with the ``wedge and integrate" pairing
\[
\langle \alpha, \beta\rangle = \int \langle \alpha, \beta\rangle_\fg .
\]
It behaves like a ``shifted symplectic'' pairing as it has cohomological degree~$-1$,
and hence it determines a degree~1 Poisson bracket $\{-,-\}$ on the graded algebra of observables.
Gauge consistency at the classical level is equivalent to the {\em classical master equation}:
\[
\{S_{mix}, S_{mix}\} = 0 .
\]

Putatively, the pairing also determines a second-order differential operator $\Delta_{BV}$ on the algebra of observables by the condition that
\[
\Delta_{BV}(FG) = (\Delta_{BV}F)G + (-1)^F F(\Delta_{BV}G) + \{F,G\}.
\]
Gauge consistency at the quantum level is encoded by the {\em quantum master equation} 
\[
\{S^{\hbar}_{mix}, S^{\hbar}_{mix}\} + \Delta_{BV} (S^{\hbar}_{mix}) = 0 .
\]
where $S^\hbar_{mix}$ is an $\hbar$-dependent action of the form $S^{\hbar}_{mix} = S_{mix} + O(\hbar)$, which we call a quantization if it satisfies the QME. 

The issue is that the BV Laplacian is ill-defined, at least naively, since it involves pairing distributional sections. 
We have already discussed how to renormalize the action functional. 
Below, we recall how to circumvent the issue with the BV Laplacian, in an approach taken by Costello in \cite{CosBook}. 

We define the mollified BV Laplacian 
\begin{equation}
\label{def of BV Lap}
\Delta_L = -\partial_{K_L}
\end{equation} 
and so the mollified BV bracket
\begin{equation}
\label{def of BV bracket}
\{F,G\}_L = \Delta_L(FG) - \Delta_L(F) G - (-1)^{|F|}F \Delta_L(G).
\end{equation} 
These allow us to articulate the scale $L$ quantum master equation.

\begin{dfn}
The scale~$L$ renormalized action $S[L]$ satisfies the scale~$L>0$ quantum master equation~(QME) if
\[
\hbar \Delta_{L} S[L] + \frac{1}{2}\{S[L],S[L]\}_L = 0 
\]
\end{dfn}

The scale $L$ action $S[L]$ satisfies the scale $L$ QME if and only if $S[L']$ satisfies the scale~$L'$ QME for every other scale~$L'$, see Lemma 9.2.2 in \cite{CosBook}.
Hence, we say a renormalized action satisfies the quantum master equation if its solves the scale~$L$ equation for some~$L$.

\subsection{The anomaly}

\def\tw{\widetilde{w}}

Not every effective action satisfies the renormalized QME, and in general one must construct solutions to the equation order by order in $\hbar$.
For mixed BF theory, however, the situation is much simpler. 
The only Feynman diagrams that can appear have genus $\leq 1$, hence the only possible anomaly occurs linearly in $\hbar$.

\begin{dfn}
The scale $L$ one-loop {\em anomaly} to quantization of mixed BF theory on $\CC \times \RR$ is
\[
\Theta [L] = \hbar^{-1} \left(Q I_{mix} [L] + \hbar \Delta_L I_{mix} [L] + \frac{1}{2} \{I_{mix}[L], I_{mix}[L]\}_L \right) \in \sO(\sE) .
\]
\end{dfn}

Since $I_{mix}[L]$ is defined as a sum of Feynman diagrams, so will the anomaly. A general fact about the anomaly is that it is given as a sum over wheel graphs. 
To state this precisely, we introduce the following notation. 

\begin{dfn}
\label{dfn: anomaly graph weight}
Given a one-loop graph $\gamma$ with distinguished internal edge $e$, the {\em weight as an anomaly} $\tw_{\gamma,e}(P_\epsilon^L, K_\epsilon, I_{mix})$ is the graph integral where the heat kernel $K_\epsilon$ is placed on the distinguished edge and the propagators $P_\epsilon^L$ are placed on the other internal edges.
\end{dfn}

\begin{lmm}[Corollary 16.0.5 of \cite{WG2}]\label{obslemma}
The limit $\Theta := \lim_{L \to 0} \Theta[L]$ exists and is a local cocycle of degree $+1$. 
Moreover, it is given by
\[
\lim_{\epsilon \to 0} \sum_{\substack{\Gamma \in \text{\rm Wheels}\\ e \in {\rm Edge}(\Gamma)}} \tw_{\gamma,e}(P_\epsilon^L, K_\epsilon, I_{mix}),
\]
where the sum is over wheels $\Gamma$ with two vertices and a distinguished inner edge $e$.
\end{lmm}

With this lemma in hand, we can get to work in analyzing the anomaly explicitly. 

\begin{lmm}
\label{lmm: anomaly}
For every wheel, the analytic weight $\tw^{an}_{\gamma,e}$ as an anomaly vanishes as a local functional.
Explicitly, 
\[
\lim_{L \to 0} \lim_{\epsilon \to 0} \tw^{an}_{\gamma,e}(P_\epsilon^L, K_\epsilon, I) = 0.
\]
\end{lmm}

The nontrivial part of this proof is quite similar to the proof of Proposition~\ref{prp: more than 2}.
The primary change is that the distinguished edge no longer has a length scale parameter;
instead, using the notation of that proof, we set $T_n = \epsilon$ everywhere.

\begin{proof}
Let the wheel have $n+1$ vertices.
When $n+1 \leq 2$, the weight vanishes because it is not possible to obtain a top form to integrate over, 
no matter what fields are inserted in the external legs.
The argument from Lemma~\ref{lmm: less than 3} applies with the minor modification that $K_\epsilon$ labels one internal edge, rather than~$P_\epsilon^L$.

We now consider the case $n+1 \geq 3$.
The graph $\gamma$, its distinguished edge $e$, and the interaction $I$ are fixed throughout the argument, 
but we wish to track the dependence on $L$ and $\epsilon$ explicitly,
so we use the notation $\tw^{an}[\epsilon,L]$ as shorthand for $w^{an}_{\gamma,e}(P_\epsilon^L, K_\epsilon, I)$.

To write the integral kernel for $\tw^{an}[\epsilon,L]$, we mimic our approach in the proof of Proposition~\ref{prp: more than 2}. 
Every graph integral $\tw^{an}_\gamma(P_\epsilon^L,I)$ determines a distribution taking as input any compactly-supported function on $(\CC \times \RR)^{n+1}$, as follows.
Label the vertices in consecutive order from $0$ to $n$ as we go around the wheel,
with the distinguished edge going from $n$ to $0$.
Each vertex is assigned some chiral interaction term,
but its details are not so important.
By definition, it will take an input function $f(p_j)$ and apply some constant-coefficient differential operator built from $\partial/\partial w_j$ to $f$; 
it will apply other such holomorphic differential operators to the incoming propagators or the heat kernel;
and then it will multiply these outputs together.
Without loss of generality, we can view the action on $f$ as absorbed into the choice of an input;
what matters is that the propagators may be acted upon by constant-coefficient holomorphic differential operators.

It is convenient to change to the coordinates $q_j = p_{j+1} - p_j$, where $j$ runs from $0$ to $n-1$.
We use $q_j = (z_j, t_j) \in \CC \times \RR$ and we use $\d q_j$ to represent the volume form $\dz_j \, \dzbar_j \, \d t_j$.
(We set $q_n = p_n$ but it plays no important role in our computation.)
In these $q$-coordinates our integral kernel is
\[
\tw[\epsilon,L](\phi) = \int_{\vec{T} \in [\epsilon,L]^{n}} \d^{n} T \, \int_{\vec{q} \in (\CC \times \RR)^{n+1}} \d^{n+1} q \,  \phi(\vec{q}) \prod_{j=0}^{n-1} D'_j E_{T_j}(q_j) \left( D'_n K_{\epsilon}^{an}(\sum_{j=0}^{n-1} q_j)\right)
\]
where $\vec{T} = (T_0,\ldots,T_{n-1})$ encodes the ``length scale'' parameter $T_j$ for the $j$th propagator and where the $D'_j$ are the $D_j$ re-expressed in the $q$ coordinates.
In light of equation~\eqref{d_z of P}, 
we see that the $D'_j$ are simply constant-coefficient differential operators built from~$\partial/\partial z_j$.

Without loss of generality, we may assume that each $D'_j$ is a monomial differential operator of the form $(\partial/\partial z_j)^{k_j}$,
as an arbitrary $D'_j$ can be written as a sum of such.
By equation~\eqref{d_z of P} and Corollary~\ref{clever diff op}, we know that 
\[
\left(\frac{\partial}{\partial z_j}\right)^{k_j} E_{T_j}(q_j)  
= \left( -\frac{\zbar_j}{4T_j}\right)^{k_j} E_{T_j}(q_j) 
= \left(  \frac{\partial}{\partial z_j} - \zeta\right)^{k_j} E_{T_j}(q_j).
\]
We also know from~\eqref{rel of G and E} that
\[
E_{T_j}(q_j) = \lambda_j G_{T_j}(q_j) = \left( c_1 \dzbar \frac{\partial}{\partial t_j} + c_2 \d t \frac{\partial}{\partial z_j} \right) G_{T_j}(q_j)
\]
with $G_T(q)$ from~\eqref{def of G} and $\lambda_j$ a constant-coefficient differential operator.
Hence
\[
D'_j E_{T_j}(q_j)  = \left(  \frac{\partial}{\partial z_j} - \zeta\right)^{k_j} \lambda_j G_{T_j}(q_j).
\]
Note that $\lambda_j$ and $\partial/\partial z_j - \zeta$ commute.

What changes, relative to the situation of Proposition~\ref{prp: more than 2}, is due to the distinguished edge,
which has fixed length scale $\epsilon$ and does not involve an integral over $T_n$.
{\it We emphasize that we are setting $T_n = \epsilon$ here in the formula for $\zeta$ from Lemma~\ref{lmm: zeta}.}

Another important change arises from the integrand contributed by the distinguished edge.
Here we have the analytic heat kernel
\[
K^{an}_\epsilon \left(\sum_{j=0}^{n-1} q_j)\right) = \frac{1}{(4 \pi T)^{3/2}} \exp\left(-\left|\sum_{j=0}^{n-1} q_j\right|^2/4T\right) \left( \sum_{j=0}^{n-1} \dz_j \, \dzbar_j \, \d t_j \right),
\]
which is also $G_\epsilon\left(\sum_{j=0}^{n-1} q_j)\right)$ multiplied by the appropriate constant 2-form, 
the sum of the $\dzbar_j \, \d t_j$.
For concision, we use $K^{(n)}_\epsilon$ to denote this kernel.
Without loss of generality, we may assume $D'_n$ is the differential operator $\zeta^{k_n}$,
because Lemma~\ref{lmm: zeta} tells us $\zeta$ acts on this edge in the same way that the $\partial/\partial z_j$ acts on the $j$th edge.

Putting these observations together and ignoring the input term $\phi$, our integrand can be written as
\[
\left(\prod_{i=0}^{n-1} \left(  \frac{\partial}{\partial z_i} - \zeta\right)^{k_i} \right) \zeta^{k_n}\left( \prod_{j=0}^{n-1} \lambda_j \right) G^{(n-1)}_{\vec{T}} K^{(n)}_\epsilon,
\]
where
\[
G^{(n-1)}_{\vec{T}} = \prod_{j=0}^{n-1} G_{T_j}(q_j).
\]
In other words, our integrand arises by applying some differential operator $D_{\vec{k}}$ to $G^{(n-1)}_{\vec{T}}K^{(n)}_\epsilon$.

What makes this description convenient is that it allows us to apply integration by parts:
we can move all the differential operators acting on $G^{(n-1)}_{\vec{T}}K^{(n)}_\epsilon$ over to the input $\phi$.
We thus find that
\[
\tw[\epsilon,L](\phi) = \pm\int_{\vec{T} \in [\epsilon,L]^{n}} \d^{n} T \, \int_{\vec{q} \in (\CC \times \RR)^{n+1}} \d^{n+1} q \, 
\left(D_{\vec{k}} \phi(\vec{q}) \right)
G^{(n-1)}_{\vec{T}}K^{(n)}_\epsilon.
\]
In short, we have reduced our graph integral to computing the expected value of $D_{\vec{k}} \phi(\vec{q})$ against a Gaussian measure depending on $\vec{T} = (T_0,\ldots,T_{n-1})$ and~$\epsilon$.

This expression has the drawback that the differential operator $D_{\vec{k}}$ explicitly depends on the length scale parameters $\vec{T}$, 
but the dependence is easy to bound.
Note that for any smooth function $f(q_0,\ldots,q_0)$, 
\[
|\zeta f| 
\leq \frac{1}{\epsilon + \sum_{j=0}^{n-1} T_j} \sum_{i = 0}^{n-1} T_i \left| \frac{\partial f}{\partial z_i}\right| 
\leq  \sum_{i = 0}^{n-1} \left| \frac{\partial f}{\partial z_i} \right|,
\]
as $T_i/(\epsilon + T_0 + \cdots + T_{n-1}) < 1$ because the $T_j$ are all positive real numbers.
In consequence, $|D_{\vec{k}} \phi(\vec{q},p_n)|$ is bounded by some compactly-supported, nonnegative function $\psi$ on $(\CC \times \RR)^{n+1}$ that is independent of~$\vec{T}$.

In short, we now know that
\[
|\tw[\epsilon,L](\phi)| \leq \int_{\vec{T} \in [\epsilon,L]^{n}} \d^{n} T \, \int_{\vec{q} \in (\CC \times \RR)^{n+1}} \d^{n+1} q \, 
\psi(\vec{q}) |G^{(n-1)}_{\vec{T}}K^{(n)}_\epsilon|
\]
Observe that $|G^{(n-1)}_{\vec{T}}K^{(n)}_\epsilon|$ is almost the Gaussian measure
\[
\frac{1}{\prod_i T_i^{\frac{3}{2}}}\exp\left( - \sum_{i=0}^{n-1} \frac{|q_i|^2}{4T_i} - \frac{1}{4\epsilon}\left|\sum_{i=0}^{n-1} q_i\right|^2\right)
\prod_{i=0}^{n-1}\d q_i,
\]
except that it is missing the $\d t_i$ and $\dzbar_i$ factors of the Lebesgue measure.
These form factors must be supplied by $\psi$ to obtain something integrable, 
which imposes conditions on which kinds of inputs matter.

If $\psi$ contributes the missing factors,
we can bound $|\tw[\epsilon,L](\phi)|$ by an integral over $(q_0,\ldots,q_{n-1})$ against this Gaussian measure.
Thus the results of Section~\ref{sec on big gaussian} kick in, with $T_n = \epsilon$,
by Taylor expanding $\psi$ and using standard formulas for moments of Gaussians.
In this case, the zeroth moment contributes a multiple of
\[
\frac{1}{ \epsilon^{3/2} T_0^{3/2} \cdots T_{n-1}^{3/2}} \left(\frac{\epsilon T_0 T_1 \cdots T_{n-1}}{\epsilon+T_0 + T_1 + \cdots + T_{n-1}}\right)^{3/2} = \frac{1}{({\epsilon+T_0 + \cdots + T_{n-1}})^{3/2}}.
\]
The integral over the $T_i$ factors then converges in the $\epsilon \to 0$ limit, as we now explain.
First, observe that 
\[
\epsilon + T_0 + \cdots + T_{n-1} >T_0 + \cdots + T_{n-1} \geq (T_0 T_1 \cdots T_{n-1})^{\frac{1}{n}}
\]
where the first inequality is obvious and the second is the arithmetic-geometric mean inequality. Hence
\begin{align*}
\int_{\vec{T} \in [\epsilon,L]^{n}} \frac{\d^{n} T}{(\epsilon +  T_0 + \cdots + T_{n-1})^{3/2}} 
&\leq \int_{\vec{T} \in [\epsilon,L]^{n}} \frac{\d^{n} T}{(T_0 \cdots T_{n-1})^{\frac{3}{2n}}}\\
&= \left( 1- \frac{3}{2n} \right) \left( \epsilon^{1-\frac{3}{2n}} - L^{1- \frac{3}{2n}} \right) .
\end{align*}
Now, we have $n \geq 2$ by hypothesis, so the right hand side converges to $-(1-3/2n)L^{1- 3/2n}$ when $\epsilon \to 0$. 
This term then goes to zero as $L \to 0$.

As at the end of the proof of Proposition~\ref{prp: more than 2}, 
the higher moments for this Gaussian also have finite $\epsilon \to 0$ limits,
and these vanish as $L$ goes to zero.
\end{proof}



\section{Chiral boundary conditions}\label{sec: boundary}

\def\RRge{\RR_{\geq 0}}
\def\RRgt{\RR_{> 0}}
\def\hata{\widehat{\frak{a}}}
\def\bHH{\overline{\HH}}
\def\Cur{{\rm Cur}}
\def\bc{{\bf c}}
\def\hotimes{{\widehat{\otimes}}}

Our gauge fixing explicitly depends on a choice of complex structure,
and hence it is natural to wonder how it interacts with the chiral WZW model on the boundary.
In this section we will begin to explore the relationship.

To start, we examine how imposing boundary conditions looks from the perspective of the the BV formalism for classical field theories.
With a convenient description in hand, 
we will show that our diagrammatic analysis from above carries over to this situation:
there is an exact one-loop quantization, without divergences or anomalies.
Finally, we discuss how we expect the factorization algebra of quantum observables to behave in this setting.

The main result of this section is the following.

\begin{thm}
Consider any chiral deformation of the mixed BF theory on $\CC \times \RR_{\geq 0}$ in the presence of the chiral boundary condition of Definition~\ref{dfn: chiral}.
Then
\begin{itemize}
\item there is no anomaly and there exists a one-loop exact quantization;
\item the chiral gauge determines a one-loop quantum correction that modifies the action by a term proportional the $\hbar$-dependent local functional
\[
\hbar \int_{\CC \times \RR_{\geq 0}} {\rm Tr}_{\fg^{ad}}(\alpha \partial \alpha) .
\]
\end{itemize}
\end{thm}

As we have seen in Corollary \ref{crl: def}, the space of quantizations of mixed BF theory is given by $\hbar \Sym^2(\fg)^\fg$ when $\fg$ is semisimple. 
Since we know there is no anomaly, the chiral gauge fixes a quantization and hence an element in $\hbar \Sym^2(\fg)^\fg$. 
Our result shows that this element is proportional to ${\rm Tr}_{\fg^{ad}}(\alpha \partial \alpha)$. 
We expect this level-shift to coincide the {\em critical level} of the semisimple Lie algebra $\fg$, but to pin down the constant would require more careful analysis. 

\subsection{A definition of the chiral boundary condition}

Our focus is on the following geometric situation. 
Let $\Sigma$ denote a Riemann surface.
Consider the closed half-space $\overline{X} = \Sigma \times \RRge$.
In essence we wish to impose a kind of Dirichlet condition: 
the $\alpha$-fields vanish on the boundary $\partial \overline{X} = \Sigma \times \{0\}$.
But to do this in the BV formalism for classical field theory requires some care.

The minor subtlety is easily seen by thinking about the analogous situation on the half-line $\RRge$.
Consider the de Rham complex
\[
\Omega^0(\RRge) \xto{\d} \Omega^1(\RRge).
\]
If we require that forms vanish on the boundary,
then we do {\em not} get a subcomplex!
Indeed, every 1-form $\omega$ (even those not vanishing on the boundary)
can appear as $\d f$ for a 0-form $f(t)$ such that $f(0) = 0$.
In short, picking a boundary condition for a differential complex requires one to impose conditions compatible with the differential.

In our case it is convenient to think about $\overline{X}$ as a half-space inside $\Sigma \times \RR$.
If we have $\alpha \in \Omega^{0,*}(\Sigma) \hotimes \Omega^*(\RR)$,
we can ask it to be even or odd with respect to reflection across $\Sigma \times \{0\}$, 
i.e., whether $\alpha$ is an eigenvector of ``time-reversal.''
Such eigenvectors are fully determined by their ``positive-time'' behavior on $\overline{X}$.
We pick out the subcomplex of odd forms:
\[
\cA_{\rm odd} = \Omega^{0,*}(\Sigma) \hotimes \left\{-1\text{-eigenspace of }\Omega^*(\RR) \right\}.
\]
Here this eigenspace means we are working with 0-forms $f(t)$ such that $f(0) = 0$ --- indeed $f^{(2k)}(0) = 0$ for every $k$ --- and with 1-forms $g(t) \d t$ such that $g^{(2k+1)}(0) = 0$ for all $k$.
(Hence $g$ satisfies a kind of Neumann boundary condition.)
Note that $\cA_{\rm odd}$ is acyclic since the -1-eigenspace is acyclic.

\begin{dfn}\label{dfn: chiral}
Given a chiral deformation $I$ of the classical mixed $BF$ theory on $\Sigma \times \RR$ with values in $\fg$,
the {\em chiral boundary condition} on $\overline{X} = \Sigma \times \RRge$ consists of restricting the $\alpha$-fields to
\[
\alpha \in \cA_{\rm odd} \otimes \fg [1]
\]
but allowing the $\beta$-fields to live in~$\cB$.
\end{dfn}

We will show in the next subsection how our computations for this field theory on $\Sigma \times \RR$ can be modified to work when the fields satisfy this chiral boundary condition.
In this sense we will see that an exact one-loop BV quantization exists in this situation.

Note that this boundary condition has an important feature:
it is a local condition along the boundary,
rather than, e.g., a spectral condition.

\begin{rmk}
We emphasize that we are imposing a boundary condition here,
which means we are picking out a ``subspace'' among all solutions to the equations of motion.
This use of the term matches the conventions in the theory of differential equations.
In the BV formalism, which involves shifted symplectic and Poisson spaces,
it is natural to require the subspace of the boundary condition to be Lagrangian or coisotropic.
There is an important alternative question
where one picks a Lagrangian foliation of the space of all solutions.
This situation falls into the BV-BFV formalism, 
which has recently received extensive investigation in~\cite{CMR1, CMR2}.
\end{rmk}

\subsubsection{An extended remark on the chiral boundary condition}

\def\sol{{\cS}{\rm ol}}

Another view on this boundary condition can be illuminating, however,
and so we quickly sketch it, 
although we do not use it in our computations below.
(This subsection is an aside rather than a full mathematical treatment.)

It is useful to recognize that for any classical field theory,
the solutions to the equations of motion satisfying some boundary condition,
which we'll denote $\sol_{\rm bc}$, are given as a fiber product:
\[
\begin{tikzcd}
\sol_{\rm bc} \arrow[r] \arrow[d] & \sol(\overline{X}) \arrow[d, "res"] \\
\cL \arrow[r, "i"] & \widehat{\sol}(\partial X)
\end{tikzcd}
\]
where 
\begin{itemize}
\item $\sol(\overline{X})$ denotes all solutions on the manifold with boundary (without any conditions imposed),
\item $\widehat{\sol}(\partial X)$ denotes all solutions on a formal neighborhood of the boundary (i.e., consider germs of solutions along the boundary and take their jets ``in the normal direction''), and
\item $\cL$ is a subspace (typically Lagrangian) inside $\widehat{\sol}(\partial X)$ (which is typically symplectic, or at least presymplectic).
\end{itemize}
In the BV formalism these spaces are derived in nature, 
because $\sol$ is encoded in a cochain complex (or, more accurately, an $L_\infty$ algebra encoding a formal moduli space),
and hence we should consider the {\em derived} fiber product.

In our discussion in the section above, we had 
\[
\widehat{\sol}(\partial X) = \Omega^{(0,*)}(\Sigma) \hotimes \widehat{\Omega^*_t} \otimes \fg[1] \oplus \Omega^{(1,*)}(\Sigma) \hotimes \widehat{\Omega^*_t} \otimes \fg,
\]
with the differential $\delbar + \d_t + \{I,-\}$ and where 
\[
\widehat{\Omega^*_t} = \CC[[t]] \xto{\d_t} \CC[[t]] \d t
\]
is the formal power series version of de Rham complex in the normal coordinate $t$.
The subspace $\cL$ consisted of the $\alpha$-fields that are even under sending $t$ to $-t$.
The {\em strict} fiber product --- where one takes the fiber product of vector spaces in each cohomological degree --- is precisely the chiral boundary condition already defined.
It is, in fact, a model of the {\em derived} fiber product because the restriction map $res$ is a levelwise surjection.

Recall that in a homological setting, we view quasi-isomorphic cochain complexes as encoding equivalent information.
More generally, in a homotopical setting, one is free to work up to weak equivalence.
In our case we have a diagram
\[
\cL \to \widehat{\sol}(\partial X) \leftarrow \sol(\overline{X})
\]
of $L_\infty$ algebras, 
and --- in place of the strict fiber product --- we are free to use any $L_\infty$ algebra quasi-isomorphic to it.

In particular, we can make a model of the derived fiber product by adjoining new fields that live just on the boundary but couple to the bulk fields.
The coupling will ensure that, at the level of cohomology, we are imposing the requirement that the $\alpha$-fields ``vanish'' on the boundary.
In practice, we use homotopical algebra as follows. 
The formal moduli spaces $\sol(\overline{X})$, $\cL$, and $\widehat{\sol}(\partial X)$ are modeled by $L_\infty$ algebras,
by the fundamental theorem of derived deformation theory.
Hence the derived fiber product is modeled by the derived fiber product of $L_\infty$ algebras.
Concretely, we take the cocone of the map 
\[
\cL \oplus \sol(\overline{X}) \xto{res - i} \widehat{\sol}(\partial X),
\]
which is models the homotopy pullback of cochain complexes, and equip it with the appropriate $L_\infty$ algebra structure.
Explicit formulas can be found (see \cite{ManFior} for details) but we do not need them here.

Note that if the boundary condition $i: \cL \to \widehat{\sol}(\partial X)$ is a map of sheaves of formal moduli problems over the manifold (i.e., is a local condition),
then taking the derived fiber product produces a stratified sheaf of formal moduli spaces over the manifold.
On the open stratum $X$ away from the boundary (the ``bulk''), one simply finds $\sol(X)$,
but on a tubular neighborhood $U$ of an open $\partial U$ in the boundary, one finds~$\widehat{\sol}(\partial U)$.
Hence local boundary conditions naturally lead to sheaves, just as solutions to the equations of motion do on manifolds without boundary.

\subsection{Why there are still no divergences and no anomalies}

We show here that even after imposing the chiral boundary condition,
there is an explicit one-loop exact BV quantization of Chern-Simons theory 
on the closed half-space $\bHH = \CC \times \RRge$.
Our construction uses the standard Euclidean metric;
we expect but do not show that our construction carries over to arbitrary manifolds of the form $\Sigma \times \RRge$ with product metrics (or even general metrics).

The approach follows closely our construction above on $\CC \times \RR$,
so we begin by writing down explicit parametrices and then show the relevant integrals have no ultraviolet divergences or (often) vanish on the nose.
Many of the key steps are essentially identical to what we have already done,
so we only discuss in any detail the aspects that change.

\subsubsection{The heat kernels and propagators on the half-space $\bHH$}

Recall the analytic piece of the heat kernel on $\CC \times \RR$:
\[
k^{an}_T(z-w,t-s) = \frac{1}{(4 \pi T)^{3/2}} e^{-|z-w|^2/4T}e^{-|t-s|^2/4T};
\]
the full kernel on the fields~is:
\[
K^{an}_T = k^{an}_T  (\dzbar - \d \overline{w}) \wedge (\dz - \d w)\wedge (\d t - \d s).
\]
The Dirichlet version, which acts on the $\alpha$-fields with chiral boundary condition, is
\begin{align*}
K^{\chi}_T(z-w,t-s) = k^{an}_T&(z-w,t-s) (\dzbar - \d \overline{w}) \wedge (\dz - \d w)\wedge (\d t - \d s) \\
&- k^{an}_T(z-w,t+s) (\dzbar - \d \overline{w}) \wedge (\dz - \d w)\wedge (\d t + \d s).
\end{align*}
We note that the dependence on $z$ and $w$ in these heat kernels is identical to that in $K^{an}_T$;
we can factor out the Gaussian function and differential forms for those coordinates.
What differs is the dependence on the ``time'' parameters $t$ and $s$.
Concretely, for the Dirichlet heat kernel, we get a sum of two terms
\[
C e^{-|t-s|^2/4T} (\d t - \d s) + C' e^{-|t+s|^2/4T} (\d t + \d s).
\]
For small scale parameter $T$, the first term is concentrated near the diagonal $\{t = s\}$; 
we have seen that such terms do not contribute divergences.
For small scale parameter $T$, the second term is concentrated near $\{t = - s \}$,
but as we have $t, s \geq 0$, this means it is concentrated near the ``boundary'' $\{ t = 0 = s\}$.
(We have a Gaussian restricted to $\RRge$.)
Already we see that ultraviolet divergences could only arise from integrals over $\partial \bHH = \CC \times \{0\}$.
Similarly, an anomaly could only arise a local functional depending on the boundary values, i.e., an integral over the boundary.

The mollified propagator is
\begin{equation}
\label{def of Dirichlet P}
\widetilde{P}_\epsilon^L (z,t; w, s) = \int_{T=\epsilon}^L \d T \, Q^* K^\chi_T(z,t; w, s) .
\end{equation}
Explicitly, ignoring the integral over $T$, 
the integrand is
\begin{equation}
\label{def of Dirichlet E}
\widetilde{E}_T(z,t;w,s) = E_T(z,t;w,s) - E^*_T(z,t;w,s).
\end{equation}
where $E_T$ is from \eqref{def of diag P} and
\begin{equation}
\label{def of Estar}
E^*_T(z,t;w,s) = \frac{e^{-(|z-w|^2+|t+s|^2)/4T} (\dz-\d w)}{(4 \pi)^{3/2}T^{5/2}} \left( (\zbar-\wbar) \,\d t +  (t+s) \,\dzbar\right)
\end{equation}
Note that $E^*_T$ has eigenvalue $-1$ with respect to the map sending $t+s$ to $-(t+s)$, i.e., it is odd. 


\subsubsection{There are no divergences or anomalies}

We use $w^{\chi}_\gamma(\widetilde{P}_\epsilon^L,I)$ to denote the weight of a graph $\gamma$ when the chiral boundary condition is imposed.
It means we use $K^\chi_T$ and integrate over the space $\bHH^{|V(\gamma)|}$, where $V(\gamma)$ is the set of vertices, where we used $K_T$ and the space $(\CC \times \RR)^{|V(\gamma)|}$ earlier.

\begin{lmm}
If a wheel $\gamma$ has one vertex, then the analytic weight $w^{\chi an}_\gamma(\widetilde{P}_\epsilon^L,I)$ vanishes for all $0 < \epsilon < L < \infty$.
In particular, the $\epsilon \to 0$ limit is zero.
\end{lmm}

This lemma holds for the same reason as Lemma~\ref{lmm: less than 3}:
the propagator $P_\epsilon^L$ is evaluated along the diagonal,
where it vanishes, and so the graph integral vanishes (even at finite $\epsilon$ and $L$). 

\begin{lmm}
If a wheel has two vertices, then the analytic weight $w^{\chi an}_\gamma(\widetilde{P}_\epsilon^L,I)$ has a well-defined $\epsilon \to 0$ limit as a distribution: 
\[
\left( \lim_{\epsilon \to 0} w^{\chi an}_\gamma(\widetilde{P}_\epsilon^L,I) \right)(\alpha) = c^{an} \int_\CC \left(\alpha|_{t = 0} \right)\wedge \partial\left( \alpha|_{t = 0}\right)
\]
where $c^{an}$ is a constant of purely analytical nature.
\end{lmm}

\begin{proof}
We let $(z,t)$ label the coordinate on the ``first'' vertex and $(w,s)$ label the coordinate on the second.
We view the graph as directed, with one edge going from the first to second vertex and the other edge going the other way.
Adding a direction breaks the symmetry of the propagator.
As a directed edge, we view one end as receiving an $\alpha$-field and the other as receiving a $\beta$-field. 
Consider the edge going from the first to the second vertex,
with the $\alpha$-field end on the first vertex.
On it we only keep the copy of the holomorphic 1-form $\d z$ and drop the term $\d w$ from the propagator:
an input $\alpha(z,t) \in \Omega^{0,*}(\CC_z) \hotimes \Omega^*(\RRge)$ wedges with $\d z$ to produce something that can be integrated over the factor~$\CC_z$.
For instance, modifying \eqref{def of Estar}, we use 
\[
E^*_T = \frac{e^{-(|z-w|^2+|t+s|^2)/4T}}{(4 \pi)^{3/2}T^{5/2}} \dz \left( (\zbar-\wbar) \,\d t +  (t+s) \,\dzbar\right)
\]
If one takes the product of these two propagators $\widetilde{E}_T(z,t; w,s) \widetilde{E}_T( w,s; z,t)$, 
we obtain 
\[
C(\zbar - \wbar) \frac{e^{-\frac{|z-w|^2}{4}\left( \frac{1}{T_0} + \frac{1}{T_1}\right)}}{T_0^{5/2} T_1^{5/2}} \d z \, \d w \, \d(\zbar - \wbar) 
\cdot e^{-\frac{(t-s)^2}{4T_0} - \frac{(t+s)^2}{4T_1}}2(t\, \d s - s \,\d t)
\]
with $C$ some constant.
(The peculiar factor $t\, \d s - s \,\d t$ appears because only the mixed products $E_T E^*_T$ contribute and they provide some cancellations.)
This term is also equal~to
\[
-C\left( \frac{1}{4T_0} + \frac{1}{4T_1}\right)^{-1}\frac{\partial}{\partial z} \left( \frac{e^{-\frac{|z-w|^2}{4}\left( \frac{1}{T_0} + \frac{1}{T_1}\right)}}{T_0^{5/2} T_1^{5/2}} \d z \, \d w \, \d(\zbar - \wbar) 
\cdot e^{-\frac{(t-s)^2}{4T_0} - \frac{(t+s)^2}{4T_1}}2(t\, \d s - s \,\d t) \right),
\]
which is convenient for applying integration by parts, as we do below.

The graph integral itself is defined by integrating the product of propagators against an input, 
namely a product $\alpha(z,t) \alpha(w,s)$,
where $\alpha \in \Omega^{0,*}_c(\CC_z) \hotimes \Omega^*_c(\RRge)$.
If one uses the final version from above and applies integration by parts,
the graph integral is
\begin{align*}
C\int_{(T_0,T_1) \in [\epsilon,L]^2} &\d T_0 \d T_1 \int_{\bHH^2}
\frac{\partial }{\partial z}(\alpha(z,t)) \alpha(w,s) 
\left( \frac{1}{4T_0} + \frac{1}{4T_1}\right)^{-1} \\
&\times\left( \frac{e^{-\frac{|z-w|^2}{4} \left(\frac{1}{T_0} + \frac{1}{T_1}\right)}}{T_0^{5/2} T_1^{5/2}} \d z \, \d w \, \d(\zbar - \wbar)
\, e^{-\frac{(t-s)^2}{4T_0} - \frac{(t+s)^2}{4T_1}} 2(t\, \d s - s \,\d t) \right).
\end{align*}
This version is messy but the $\epsilon \to 0$ limit is simple,
as we now show.

We wish to characterize the distribution encoded by this graph integral, 
so it suffices to consider inputs $\alpha$ that factor as $a(z) f(t)$,
as such products are dense among all $\alpha$-fields.
For such products, the integral over $\bHH^2$ also factors:
\begin{align*}
\int_{(z,w) \in \CC^2} & \frac{\partial }{\partial z}(a(z))a(w) 
e^{-\frac{|z-w|^2}{4} \left(\frac{1}{T_0} + \frac{1}{T_1}\right)} \d z \, \d w \, \d(\zbar - \wbar) \\
&\times \int_{(t,s) \in \RRge^2} f(t)f(s) e^{-\frac{(t-s)^2}{4T_0} - \frac{(t+s)^2}{4T_1}} 2(t\, \d s - s \,\d t).
\end{align*}
We denote the first term --- given by the integral over $\CC^2$ --- by $I_\CC(\epsilon,L)$,
and we denote the second term --- given by the integral over $\RRge^2$ --- by~$I_{\RRge}(\epsilon,L)$.

The factor $I_\CC(\epsilon,L)$  can be tackled by Wick's lemma,
in the style of several proofs above.
The zeroth moment is precisely the desired integral $\int_\CC a\partial a$, up to a constant times~$(1/T_0 + 1/T_1)^{-1}$.
Higher moments will contribute further powers of $(1/T_0 + 1/T_1)^{-1}$, which only improve the $\epsilon \to 0$ limit for the integral over the scale parameters $T_0$ and~$T_1$.

The second factor $I_{\RRge}(\epsilon,L)$ requires arguments different from what we've done so far, 
and we will merely obtain convenient bounds.
First, note that $f$ is a compactly-supported differential form on $\RRge$ that is the restriction of a form on the whole line $\RR$ that is a $(-1)$-eigenvector under reflection (i.e., an ``odd'' form). 
Hence the 0-form component $f_0(t)$ of $f$ is an odd function --- so $f_0(t)$ vanishes to first order at the origin ---
while in the 1-form component $f_1(t) \d t$, 
the function $f_1(t)$ is even.
We see then that
\begin{align*}
I_{\RRge}(\epsilon,L) 
&= \int_{(t,s) \in \RRge^2}  f(t)f(s) e^{-\frac{(t-s)^2}{4T_0} - \frac{(t+s)^2}{4T_1}} (t\, \d s - s \,\d t)\\
&= \int_{(t,s) \in \RRge^2} \left( t f_1(t)f_0(s) - s f_0(t) f_1(s)\right) e^{-\frac{(t-s)^2}{4T_0} - \frac{(t+s)^2}{4T_1}} \, \d t\,\d s\\
&= \int_{(t,s) \in \RRge^2} ts \phi(t,s) e^{-\frac{(t-s)^2}{4T_0} - \frac{(t+s)^2}{4T_1}} \, \d t\,\d s
\end{align*}
where $\phi$ is a compactly-supported function on $\RRge^2$ such that
\[
ts \phi(t,s) = t f_1(t)f_0(s) - s f_0(t) f_1(s).
\]
Such a $\phi$ exists due to the boundary behavior of the $f_j$.
In consequence we find a simple bound
\[
|I_{\RRge}(\epsilon,L)| \leq \max_{(t,s) \in\RRge^2} |\phi(s,t)| 
\int_{(t,s) \in \RRge^2} ts  e^{-\frac{(t-s)^2}{4T_0} - \frac{(t+s)^2}{4T_1}} \, \d t\,\d s,
\]
as the absolute value of the integral is bounded by the integral of the absolute value of the integrand.

The right hand side is easy to bound, thankfully.
Let us change coordinates to $u = t+s$ and $v = t-s$. 
Then 
\[
\int_{(t,s) \in \RRge^2} ts  e^{-\frac{(t-s)^2}{4T_0} - \frac{(t+s)^2}{4T_1}} \, \d t\,\d s = \frac{1}{8} \int_{v = 0}^\infty \int_{u=-v}^v (v^2 - u^2) e^{-\frac{v^2}{4T_0} - \frac{u^2}{4T_1}}\,\d u \, \d v.
\]
Apply the triangle inequality:
\begin{align*}
\left| \frac{1}{8} \int_{v = 0}^\infty \int_{u=-v}^v (v^2 - u^2) e^{-\frac{v^2}{4T_0} - \frac{u^2}{4T_1}}\,\d u \, \d v \right|
& \leq \frac{1}{8} \int_{v = 0}^\infty \int_{u=-v}^v (v^2 + u^2) e^{-\frac{v^2}{4T_0} - \frac{u^2}{4T_1}}\,\d u \, \d v \\
& \leq \frac{1}{8} \int_{v = 0}^\infty \int_{u=-\infty}^\infty (v^2 + u^2) e^{-\frac{v^2}{4T_0} - \frac{u^2}{4T_1}}\,\d u \, \d v,
\end{align*}
since extending the domain of integration merely increases the value.
But now we can apply Wick's lemma directly:
\begin{align*}
\left| \frac{1}{8} \int_{v = 0}^\infty \int_{u=-v}^v (v^2 - u^2) e^{-\frac{v^2}{4T_0} - \frac{u^2}{4T_1}}\,\d u \, \d v \right|
&\leq \frac{\sqrt{\pi}}{8 } \int_{v = 0}^\infty (T_0^{1/2} v^2 +  T_0^{3/2}/2) e^{-\frac{v^2}{4T_0} }\, \d v\\
&= \frac{\pi}{8 } \left( \frac{T_0^{1/2} T_1^{3/2}}{4} +   \frac{T_0^{3/2}T_1^{1/2}}{2} \right) \\
&\leq \frac{\pi}{16} T_0^{1/2} T_1^{1/2} \left( T_0 + T_1\right).
\end{align*}
We do not care about the constant but the exponents of the scale parameters $T_0$ and $T_1$ are crucial. 

Putting our observations together, we see that for an input $\alpha$ of the form $a(z) f(t)$, we have a bound on the graph integral
\begin{align*}
|w^{\chi an}_\gamma(\widetilde{P}_\epsilon^L,I)(\alpha)| 
\leq 
C(f) \int_\CC a\partial a &\int_{(T_0,T_1) \in [\epsilon,L]^2} \,\d T_0 \d T_1 
\left( \frac{1}{T_0} + \frac{1}{T_1}\right)^{-2} \frac{\left( T_0 + T_1\right)}{T_0^{2} T_1^{2}} \\
&+ \text{terms involving higher moments in $a$},
\end{align*}
where $C(f)$ is a constant depending only on $f$ (essentially the maximum of $|\phi|$)
and where the corrections depend on the higher moments of the integral over $a$. 
These corrections have stronger convergence as $\epsilon \to 0$, 
just as in the argument appearing at end of the proof of Proposition~\ref{prp: more than 2}. 

Observe that
\[
\left( \frac{1}{T_0} + \frac{1}{T_1}\right)^{-2} = \frac{ T_0^{2} T_1^{2}}{(T_0 + T_1)^{2}}
\]
so we need to compute the easy integral
\[
\int_{(T_0,T_1) \in [\epsilon,L]^2}  \frac{\d T_0 \,\d T_1}{T_0 + T_1}.
\]
We obtain
\[
L \log(2L) - \epsilon \log(L + \epsilon) - L \log(\epsilon + L) + \epsilon \log(2\epsilon)
\]
which has a finite and nonzero limit as $\epsilon \to 0$.
\end{proof}

\begin{prp}
If a wheel $\gamma$ has more than two vertices, then the analytic weight $w^{\chi an}_\gamma(\widetilde{P}_\epsilon^L,I)$ has a well-defined $\epsilon \to 0$ limit as a distribution.
\end{prp}

\begin{proof}
Let the number of vertices $|V(\gamma)|$ be $n+1$.
To prove this proposition, we will show that for any compactly-supported input $\phi$ satisfying the chiral boundary condition
(which is given by choosing $n+1$ fields of $\alpha$ type),
the absolute value of the weight $|w^{\chi an}_\gamma(\widetilde{P}_\epsilon^L,I)(\phi)|$ is bounded above by an integral whose dependence on $\epsilon$ converges as it goes to zero.
The structure of the problem is quite similar to that in Proposition~\ref{prp: more than 2} and so we discuss only what is different from the proof there.

Note that each propagator contributes two terms, 
one that is the propagator on $\CC \times \RR$ restricted to $\bHH$ and 
one that involves the sum $s+t$ of the ``time'' parameters rather than the difference.
Schematically, we have $\widetilde{P} = P + P^*$, where $P$ denotes the ``original'' propagator of~\eqref{def of P} and $P^*$ denotes the term associated to the integral over $E^*_T$.
Expanding out the product of the propagators, we find a term $P(p_0 - p_1) \cdots P(p_n - p_0)$ that is purely a product of the first type.
We have already show that integration against this term has a well-defined $\epsilon \to 0$ limit.

Hence we only need to consider the mixed products, where at least one factor is of the form $P^*(p_k; p_{k+1})$ 
where $p_k = (z_k, t_k) \in \CC \times \RRge$.
In a mixed product of propagators, if the $k$th edge contributes the original propagator $P$ as a factor, 
the integral is concentrated near $t_k = t_{k+1}$ for small values of $T$.
If the $k$th edge contributes the new propagator $P^*$ as a factor, 
the integral is concentrated near $t_k + t_{k+1} = 0$ for small values of $T$, 
which means region where both $t_k$ and $t_{k+1}$ are close to 0.
Hence in a mixed product of propagators, the whole integral is concentrated in the region where $t_j \approx 0$ for all time parameters~$t_j$.

As the integrands over the complex coordinates $z_k$ remain unchanged, 
we focus on the integrals over the time parameters~$t_k$.
Here there's a nice interplay between the boundary condition and the boundary behavior of $E^*_T$.
The inputs are $\alpha$-fields and hence they satisfy the boundary condition.
As we'll now show, this condition assures us that the integral over $\bHH^{n+1}$, where $n+1$ is the number of vertices of $\gamma$, 
leaves an integral over the time parameters $T_0, \ldots, T_n \in [\epsilon, L]$ that is convergent as $\epsilon \to 0$.
Explicitly, we will see that we can apply the AM-GM inequality argument to the integrand over the parameters $T_k$ that we used at the end of the proof of Proposition~\ref{prp: more than 2}.

Suppose the $k$th input $\alpha_k$ is a 1-form with respect to $t_k$, 
and so it factors as $\alpha_k = f(t_k) \d t_k$ and $f$ is an even function respect to $t_k$.
On the other hand, when wedging this input with the propagators, 
only the factors without a $\d t_k$ term contribute. 
Any term from an integrand $\widetilde{E}_T$ without a $\d t_k$ factor is weighted by a copy of $t_k$ (more accurately $t_k -t_{k+1}$ or $t_k + t_{k+1}$),
so that the overall graph integral vanishes to order one (at least) as $t_k \to 0$.
Applying Wick's lemma for Gaussians on a half-line, 
we need to compute first and higher order moments.
These contribute at least one power of~$T_k$.
The $\widetilde{E}_T$ already contains a factor $T_k^{-5/2}$, 
so we get an overall factor of $T_k^{-3/2}$,
which is enough for the argument by the AM-GM inequality.
 
Suppose instead that the $k$th input $\alpha_k$ is a 0-form with respect to $t_k$, 
and so it is an odd function respect to $t_k$ and hence vanishes to order one (at least) as $t_k \to 0$.
When wedging this input with the propagators, 
only the factors with a $\d t_k$ term contribute. 
Hence the overall graph integral vanishes to order one (at least) as $t_k \to 0$.
Applying Wick's lemma for Gaussians on a half-line, 
we need to compute first and higher order moments.
These contribute at least one power of~$T_k$.
Again, this suffices to apply the argument by the AM-GM inequality.
\end{proof}

This argument extends, with tiny modifications, to the computation of the anomaly.
Let $\widetilde{w}^\chi_{\gamma,e}(\widetilde{P}_\epsilon^L, K_\epsilon, I)$ denote the weight of a graph $\gamma$ with a distinguished internal edge $e$ when we have imposed the chiral boundary condition (cf. Definition~\ref{dfn: anomaly graph weight}).

\begin{lmm}
For every wheel, the analytic weight $\tw^{\chi an}_{\gamma,e}$ as an anomaly vanishes as a local functional.
Explicitly, 
\[
\lim_{L \to 0} \lim_{\epsilon \to 0} \tw^{\chi an}_{\gamma,e}(\widetilde{P}_\epsilon^L, K^\chi_\epsilon, I) = 0.
\]
\end{lmm}

\begin{proof}
As in the preceding proof, the novel aspect --- relative to the proof of Lemma~\label{lmm: anomaly} --- is that our propagators and heat kernel each consist of two terms. 
One term is the restriction of the propagator or kernel we used on $\CC \times \RR$ to $\bHH$,
while the other term depends on the sum $s+t$ of time parameters rather than their difference $s-t$. 
As in the preceding proof, we find that this second term means that as its scaling parameter $T_k$ gets small,
the integral is concentrated near the boundary, with all time parameters $t_j \approx 0$.
The chiral boundary condition assures us that the overall integrand vanishes at least to order one as these time parameters approach zero.
Wick's lemma for the Gaussian on a half-line assures us that the integral over $\bHH^{n+1}$ leaves an integral over the scale parameters $T_0,\ldots, T_n \in [\epsilon,L]$ that can be bounded by an AM-GM inequality argument.
This bound ensures that an $\epsilon \to 0$ limit exists, and that the $L \to 0$ limit is actually zero,
just as we saw in the proof of Lemma~\ref{lmm: anomaly}.
\end{proof}

\subsection{The expected result at the level of factorization algebras}

Above we explained why and how the Feynman diagrams are well-behaved on the manifold $\CC \times \RR$ by using explicit heat kernels.
Similar arguments should apply to a manifold of the form $\Sigma \times \RR$, 
where $\Sigma$ is a Riemann surface.
This style of construction is global in nature,
so it is natural to ask about the local story,
particularly the operator products of this quantum field theory.

In the claim below we will give a precise description of these operator products using the language of factorization algebras.
This language encodes systematically the operators of a field theory, whether classical or quantum, and organizes them by their support on the spacetime manifold.
For example,
in Chapter 5, \S4 of \cite{CG1}, a factorization algebra $\cF^\kappa$ on Riemann surfaces is constructed for each Lie algebra $\g$ with invariant symmetric pairing $\kappa$.
The key theorem is that $\cF^\kappa$ recovers the affine Kac-Moody vertex algebra $V_\kappa$ for $\g$ with the pairing $\kappa$ when one examines the structure maps of $\cF^\kappa$ for inclusions of disks into larger disks.
As another example, Chapter 4, \S5 of \cite{CG1} constructs and examines the factorization algebra on oriented 3-dimensional manifolds arising from abelian Chern-Simons theory.
In \cite{CFG} there is an exploration of the factorization algebra $\Obs^q_{CS}$ of nonabelian Chern-Simons theory;
but its existence is guaranteed by the main theorem of~\cite{CG2}.
Our work in this paper suggests how to relate the factorization algebra $\cF^\kappa$, which lives on Riemann surfaces, 
to $\Obs^q_{CS}$, which lives on oriented, smooth 3-manifolds.

Our focus is on the following geometric situation. 
Let $\Sigma$ denote a Riemann surface.
Consider the closed half-space $\overline{X} = \Sigma \times \RRge$,
let $X$ denote the open half-space $\Sigma \times \RRgt$,
and let $\pi: \overline{X} \to \CC$ denote the projection map.
We will sketch how to construct a factorization algebra $\Obs^\q_\alpha$ for abelian Chern-Simons theory on $\overline{X}$ with the boundary condition that the $\alpha$-fields vanish on the boundary $\partial \overline{X} = \Sigma \times \{0\}$.

\begin{clm}
\label{claim on fact alg}
The BV quantization of Chern-Simons theory on $\overline{X}$ with chiral boundary condition on its boundary $\Sigma$ yields a factorization algebra $\Obs^\q_\alpha$ on $\overline{X}$ that is stratified in the sense that
\begin{itemize}
\item on the interior $X$, there is a natural quasi-isomorphism 
\[
\Obs^\q_{CS} \simeq \Obs^\q_\alpha|_X,
\] 
and
\item for any connected, open neighborhood $U$ of the boundary $\partial \overline{X} = \Sigma \times \{0\}$, there is a natural quasi-isomorphism 
\[
\cF^\kappa \simeq (\pi_U)_*\Obs^\q_\alpha
\] 
of factorization algebra on $\Sigma$, where the level $\kappa$ is determined by the quantization.
\end{itemize}
\end{clm}

\begin{rmk}
We use the term ``claim'' here because we merely sketch how arguments from \cite{CosBook, CG2} should extend to this situation.
A proof would require tools and arguments that work, in fact, at a higher and more convenient level of generality 
--- leading to analogs of the main results of \cite{CG2} --- 
and we anticipate such a treatment appearing in the near future~\cite{Eugene}.
Our computations in this paper are compatible with those constructions and will imply the claim as a consequence of that general theory.
On the other hand, for {\em abelian} Chern-Simons theory,
we give, with Rabinovich, a complete treatment that avoids any Feynman diagrams in a companion paper~\cite{CSWZW}.
\end{rmk}

This factorization algebra would thus exhibit the desired phenomenon, 
as it is precisely the  Chern-Simons system in the ``bulk'' $X$ but becomes the chiral currents on the boundary.

Let us sketch now how the arguments from \cite{CosBook, CG2} extend to our situation.
The key observation in \cite{CosBook} --- for the construction of factorization algebras --- is that the analysis underlying perturbative BV quantization is quite flexible:
one is free to work with any parametrices of the generalized Laplacian $[Q,Q^*]$ arising from a choice of gauge-fixing $Q^*$,
and there is an explicit isomorphism between constructions using one parametrix (e.g., arising from a heat kernel construction) and another.
In particular, one produces the interaction term $I[\Phi]$ for each parametrix $\Phi$,
generalizing the interactions $I[L]$ we described on $\CC \times \RR$.
(Indeed, $L$ can be seen as labeling the parametrix $\int_0^L K_t \, \d t$.)
Similarly, just as for an interaction term $I$, every observable $O$ has a representative $O[\Phi]$ for each parametrix.
A central result of \cite{CG2} is that the cochain complex of observables for parametrix $\Phi$ is isomorphic to the cochain complex of observables for any other parametrix.

One can always pick a parametrix whose support is arbitrarily close to the diagonal.
In this way one can give a precise notion of the ``support'' of an observable $O$,
as an open set $U$ such that $O[\Phi] \subset U$ for $\Phi$ with support sufficiently close to the diagonal.
Observables with disjoint support can be multiplied in the naive way, 
i.e., by multiplying $O[\Phi]$ with $O'[\Phi]$ so long as $\Phi$ is small enough.
This multiplication provides the factorization structure.

The arguments justifying these observations carry over to nice boundary conditions for elliptic complexes,
such as the one we study here.
Pinning down ``nice'' at convenient level of generality, so that the arguments {\em do} carry over, is the challenge here;
to get a factorization algebra, one certainly wants the boundary condition to be local along the boundary (as opposed to, e.g., a spectral boundary condition).
As we have shown how to quantize using the parametrices that arise from the heat kernel appropriate to this boundary condition,
our results imply the existence of interaction terms and observables for other parametrices.
It is, for instance, easy to see that if one uses the ``fake heat kernels'' of~\cite{CosBook},
which simply multiply the heat kernel we use here by a smooth bump function with support along the diagonal,
our computations here assure that no divergences arise.
In particular, one can use parametrices with support in a small neighborhood of the diagonal,
and hence one obtains a factorization algebra.

\bibliographystyle{alpha}  
\bibliography{cs}

\end{document}